\documentclass[pre,aps,floatfix,10pt,superscriptaddress]{revtex4-1}
\usepackage[english]{babel}
\usepackage[T1]{fontenc}
\usepackage[latin9]{inputenc}
\usepackage{amsmath,amssymb}
\usepackage[svgnames]{xcolor}
\usepackage{graphicx}
\usepackage{subcaption}
\graphicspath{{./figures/}}

\newcommand{\set}[1]{\ensuremath{\mathcal{#1}}}
\newcommand{\obs}{\ensuremath{A}}

\newcommand{\nsample}{\ensuremath{N}}
\newcommand{\niter}{\ensuremath{J}}
\newcommand{\nexp}{\ensuremath{K}}

\usepackage{acro}
\DeclareAcronym{ams}{
  short=AMS,
  long=Adaptive Multilevel Splitting,
  long-format=\itshape,
}
\DeclareAcronym{tams}{
  short=TAMS,
  long=Trajectory-Adaptive Multilevel Sampling,
  long-format=\itshape,
}

\DeclareAcronym{gktl}{
  short=GKTL,
  long=Giardina-Kurchan-Tailleur-Lecomte,
}
\DeclareAcronym{ou}{
  short=OU,
  long=Ornstein-Uhlenbeck,
}
\DeclareAcronym{pdf}{
  short=PDF,
  long=Probability Density Function,
}
\DeclareAcronym{cdf}{
  short=CDF,
  long=cumulative distribution function,
}

\newtheorem{theo}{Theorem}

\usepackage[colorlinks=true,linkcolor=DarkBlue,urlcolor=DarkBlue,citecolor=FireBrick]{hyperref}

\begin{document}

\title{Computing return times or return periods with rare event algorithms}

\author{Thibault Lestang}
\email{thibault.lestang@ens-lyon.fr}
\affiliation{Univ Lyon, ENS de Lyon, Univ Claude Bernard, CNRS, Laboratoire de Physique, F-69342 Lyon, France}
\affiliation{Univ Lyon, Ecole Centrale de Lyon, Univ Claude Bernard, CNRS, Laboratoire de M\'ecanique des Fluides et d'Acoustique, F-69134 Ecully cedex, France}
\author{Francesco Ragone}
\email{francesco.ragone@unimib.it}
\affiliation{Department of Earth and Environmental Sciences,
University of Milano-Bicocca, Milan, Italy}
\author{Charles-Edouard Br{\'e}hier}
\email{brehier@math.univ-lyon1.fr}
\affiliation{Univ Lyon, Universit\'e Claude Bernard Lyon 1, CNRS UMR 5208, Institut Camille Jordan, 43 blvd. du 11 novembre 1918, F-69622 Villeurbanne cedex, France}
\author{Corentin Herbert}
\email{corentin.herbert@ens-lyon.fr}
\affiliation{Univ Lyon, ENS de Lyon, Univ Claude Bernard, CNRS, Laboratoire de Physique, F-69342 Lyon, France}
\author{Freddy Bouchet}
\email{freddy.bouchet@ens-lyon.fr}
\affiliation{Univ Lyon, ENS de Lyon, Univ Claude Bernard, CNRS, Laboratoire de Physique, F-69342 Lyon, France}

\begin{abstract}
  The average time between two occurrences of the same event, referred to as its \emph{return time} (or return period), is a useful statistical concept for practical applications.
  For instance insurances or public agency may be interested by the return time of a $10$ m flood of the Seine river in Paris.
  However, due to their scarcity, reliably estimating return times for rare events is very difficult using either observational data or direct numerical simulations.
  For rare events, an estimator for return times can be built from the extrema of the observable on trajectory blocks.
  Here, we show that this estimator can be improved to remain accurate for return times of the order of the block size.
  More importantly, we show that this approach can be generalised to estimate return times from numerical algorithms specifically designed to sample rare events.
  So far those algorithms often compute probabilities, rather than return times.
  The approach we propose provides a computationally extremely efficient way to estimate numerically the return times of rare events for a dynamical system, gaining several orders of magnitude of computational costs.
  We illustrate the method on two kinds of observables, instantaneous and time-averaged, using two different rare event algorithms, for a simple stochastic process, the Ornstein--Uhlenbeck process.
  As an example of realistic applications to complex systems, we finally discuss extreme values of the drag on an object in a turbulent flow.
\end{abstract}


\maketitle

\tableofcontents{}

\clearpage{}

\section{Introduction}

In many physical systems, the mean state and the typical fluctuations about this state, usually studied in statistical physics, are not the only quantities of interest.
Indeed, fluctuations far away from the mean state, although they are usually very rare, can play a crucial part in the macroscopic behaviour of the system.
For instance, they can drive the system to a new metastable state, possibly with radically different properties~\cite{Kramers1940}.
Such transitions arise in a wide variety of situations, such as Josephson junctions~\cite{Kurkijarvi1972}, quantum oscillators~\cite{Dykman2012}, turbulent flows~\cite{Bouchet_Simonnet_2008}, magneto-hydrodynamics dynamos~\cite{Berhanu2007}, diffusion-controlled chemical reactions~\cite{Calef1983}, protein folding~\cite{Noe2009}, climate dynamics~\cite{Paillard1998}.
Even if the system returns to its original state after undergoing the large fluctuation, the impact of this event may be so large that it is worth being studied on its own.
One may think for instance about heat waves~\cite{Robine2008} and tropical cyclones, rogue waves in the ocean~\cite{Dysthe2008}, strong dissipative events in turbulent flows~\cite{Yeung2015}, shocks in financial markets~\cite{EmbrechtsBook}.
Here, we are concerned with the study of such atypical fluctuations starting from the equations (deterministic or stochastic) which govern the dynamics of the system.
This approach is different from and complementary to the purely statistical methods which try to extract the best possible information about the distribution of rare events from an existing timeseries, such as, for instance, extreme value statistics~\cite{Ghil2011,Fortin2015,LucariniExtremesBook}.

The theoretical framework which has been developed over the last decades in statistical physics to tackle this problem is that of \emph{large deviation theory}~\cite{FreidlinWentzellBook,EllisBook,DenHollanderBook,Touchette2009,VulpianiBook}.
Numerical methods have also been developed to efficiently sample rare events, which are not amenable to classical Monte-Carlo methods~\cite{AsmussenGlynn2007,LandauBinder2015,Liu2008}; see~\cite{Bucklew2004,RubinoTuffin2009} for general references on rare event simulation.
Those algorithms can be roughly divided into two main classes: those which work in state space, and evolve a population of \emph{clones} of the system according to selection rules biased to favour the appearance of the desired rare event~\cite{Grassberger2002,DelMoral2005,Cerou2007,Giardina2011,Rolland2016}, and those which try to sample directly in path space the histories of the system which exhibit the phenomenon of interest~\cite{Dellago2002,E2002,E2004,laurie2015computation,Grafke2015b,grigorio2017instantons}.
They can be used either for stochastic processes or deterministic chaotic dynamical systems~\cite{wouters2016rare}.
Most of those algorithms ultimately compute either one-time statistics (typically, the stationary probability distribution of the system, for which they sample efficiently the tails, or alternatively, large deviation rate functions or scale cumulant generating functions), or reactive trajectories corresponding to the transition between two metastable states.

From a modelling perspective, it is natural to assume that successive occurrences of a rare event are independent from one another~\cite{EmbrechtsBook,leadbetter_extremes_1983,doucet_sequential_2001}.
Then, the average number of events occurring in a time interval is proportional to the length of that interval.
This is the definition of a Poisson process.
In this case, all the statistics are encoded in a single parameter, the rate of the Poisson process.
In the following, we will assume that we are dealing with the simple case of a well identified process that can be described by a single return time or rate.
This is often a sufficient framework; indeed the long time behaviour of many systems can be described phenomenologically, or exactly in some limits, as Markov processes described by a set of transition rates describing independent processes (see for instance~\cite{FreidlinWentzellBook} for systems driven by a weak noise).
We note however that many other physical systems are not amenable to such a simple effective Markov processes, for instance structural glasses or amorphous media.

For many practical applications, the most useful information about a rare event is the \emph{return time}: it is the typical time between two occurrences of the same event.
This is how hydrologists measure the amplitude of floods for instance~\cite{Sveinsson2002}.
As a matter of fact, one of the motivations of Gumbel, a founding father of extreme value theory, was exactly this problem~\cite{Gumbel1941}.
Other natural hazards, such as earthquakes \cite{Corral2005} and landslides~\cite{Peres2016}, are also ranked according to their return time.
Similarly, climatologists seek to determine how the frequency of given heat waves~\cite{Meehl2004b,Rahmstorf2011} or cold spells~\cite{Cattiaux2010} evolves in a changing climate~\cite{Shepherd2016}.
Public policies rely heavily on a correct estimate of return times: for instance, in the United States, floodplains were defined in the National Flood Insurance Program in 1968 as areas vulnerable to events with a 100-year return time.
Such definitions are then used to determine insurance policies for home owners.
In the industry as well, return times are the metric used by engineers to design systems withstanding a given class of events.
Another property describing rare events of a time series is the average time between successive records~\cite{Godreche2017}; here, because of its importance in practical applications, we focus on the return time, i.e. the average time between events of a given amplitude.
Just like the extreme values of any observable, the return time of a rare event is very difficult to estimate directly from observational or numerical data, because extremely long timeseries are necessary.

The return time may be estimated heuristically by interpreting it as a \emph{first-passage time}.
The \emph{first-passage time} (sometimes also called \emph{first exit time}) is defined as the time it takes a stochastic process to reach the boundary of a given domain for the first time; the properties of this random variable have been studied extensively in statistical physics~\cite{RednerBook,Bray2013}.
Then, the return time (or return period) $r(a)$ for an event of amplitude $a$ (return level) may be at first sight related to the inverse of the stationary probability $p_s$: $r(a)=\tau_{c}(a)/{p_{s}(a)}$, where the correlation time $\tau_{c}(a)$ usually depends on $a$ but remains bounded when $p_{s}(a)$ goes to zero.
This is true for instance for a system perturbed by a small-noise $\epsilon$ at the level of large deviations: $r(a) \underset{\epsilon \rightarrow 0}{\asymp} e^{U(a)/\epsilon}$, where the quasi-potential $U$ is defined by $p_s(a) \underset{\epsilon \rightarrow 0}{\asymp} e^{-U(a)/\epsilon}$~\cite{FreidlinWentzellBook}.
However the return time is only roughly proportional to the inverse of the stationary probability~\cite{NicolisNicolis2007}.
In order to compute $\tau_{c}(a)$ one has to go beyond large deviation theory.
For instance for gradient dynamics and for first exit time problems, exact formulas exists~\cite{Langer1969,GardinerBook,RiskenBook}, valid at leading order in $\epsilon$ (we stress that different formula are obtained depending on the hypothesis made on the domain that the particle exits).
Generalisations to irreversible non gradient dynamics also exist (see~\cite{bouchet2016generalisation} and references therein).
From these computations, it appears clearly that $\tau_{c}(a)$ is not simply related to ${p_{s}(a)}$ and that the return time $r(a)$ is a trajectory property, not amenable to a one point statistics like ${p_{s}(a)}$.

There is thus a need to develop rare event algorithms specifically designed for computing return times, valid also when large deviation estimates are not relevant.
This is the aim of this paper.
The approach developed in this work relies on the combination of two observations.
First, if one assumes that rare events are described by a Poisson process, then return times can be related to the probability of observing extrema over pieces of trajectories, which are of duration much larger than the correlation time of the system, but typically much smaller than the computed return times.
Second, several classes of rare event algorithms can be easily generalised to compute the probability of extrema over pieces of trajectories, rather than to compute single point statistics.
We show that combining these two remarks enables us to build a powerful tool to compute return times in an elementary way with simple and robust algorithms.
As a side remark, we also discuss a new way to construct return time plots from a timeseries, which provides an important improvement for return times moderately larger than the sampling time, even when we are not using a rare event algorithm.

We illustrate the method by computing return times, first for an instantaneous observable (one-point statistics) using the \acf{ams} algorithm~\cite{Cerou2007,Cerou2011}, and second for a time-averaged observable, using both the \ac{ams} algorithm and the Del Moral-Garnier algorithm~\cite{DelMoral2005} (or equivalently the Giardina-Kurchan algorithm~\cite{Giardina2006} in a non-stationary context).
The computation of return times with the \ac{ams} algorithm leads us to define a generalisation called the \acf{tams} algorithm.
This generalisation has several practical advantages: it computes directly return times $r(a)$ for a full range of return level $a$ rather than a single one, and it avoids the tricky estimation of time scale on an auxiliary ensemble, and the sampling from this auxiliary ensemble.
As a test, we first carry out these computations for a simple stochastic process, the \acf{ou} process, for which analytical results are available and the accuracy and efficiency of the algorithm can be tested thoroughly.
Then, to demonstrate the usefulness of the method in realistic applications, we briefly showcase a problem involving a complex dynamical system: extreme values of the drag on an object immersed in a turbulent flow.

The structure of this paper is as follows: in section~\ref{sec:Return_Time_Plots}, we introduce the method to compute return times from a timeseries and from rare event algorithms.
We define the \acf{tams} algorithm in section~\ref{sec:AMS}.
We apply the method to compute return times for the instantaneous and time-averaged observables for an Ornstein--Uhlenbeck process, respectively, in section~\ref{sec:AMS} (using the \ac{tams} algorithm) and~\ref{sec:gktl} (using both the \ac{tams} and the \acf{gktl} algorithms).
Finally, we introduce the application to complex dynamical systems in section~\ref{sec:applications}, before presenting our conclusions in section~\ref{sec:cl}.
We discuss in the conclusions the range of applicability of these algorithms.

\section{Return Times: Definition and Sampling Methods}
  \label{sec:Return_Time_Plots}

\subsection{Computing return times from a timeseries}\label{sec:return_time_timeseries}

\subsubsection{Definition of return times}

\begin{figure}
  \centering
  \begin{subfigure}{1.0\textwidth}
    \centering
    \includegraphics[scale=0.6]{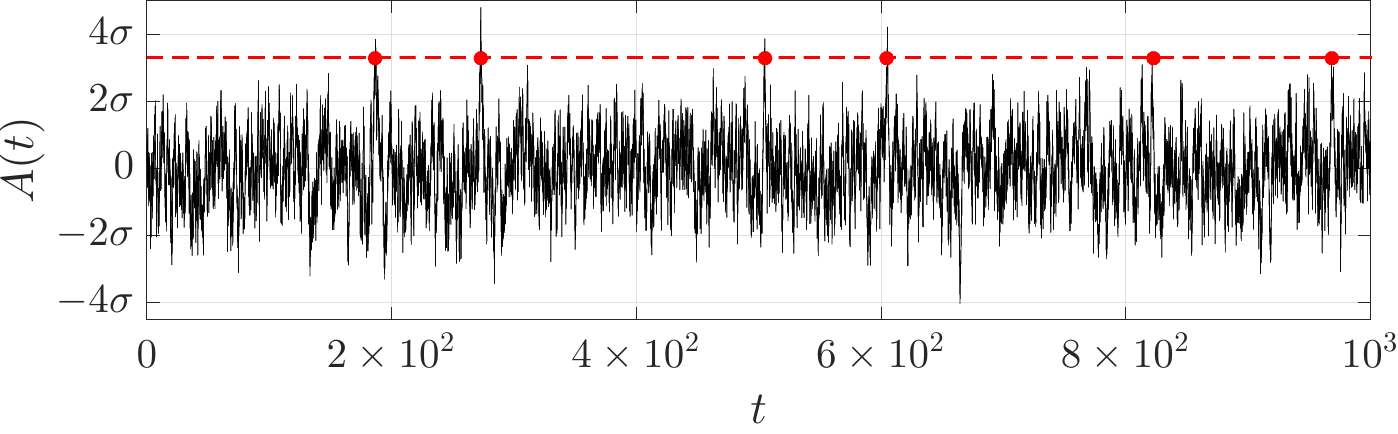}
    \caption{Sample timeseries (black curve), generated from an Ornstein--Uhlenbeck process~\eqref{eq:OUprocess} ($\alpha=1$, $\epsilon=1/2$; $\sigma=1/\sqrt{2}$ is the standard deviation).
      We are interested in fluctuations which reach a prescribed threshold $a$ (red curve). These events are identified by the red dots.}
    \label{fig:courbe_signal_avec_seuil}
  \end{subfigure}
  \begin{subfigure}{1.0\textwidth}
    \centering
    \includegraphics[scale=0.6]{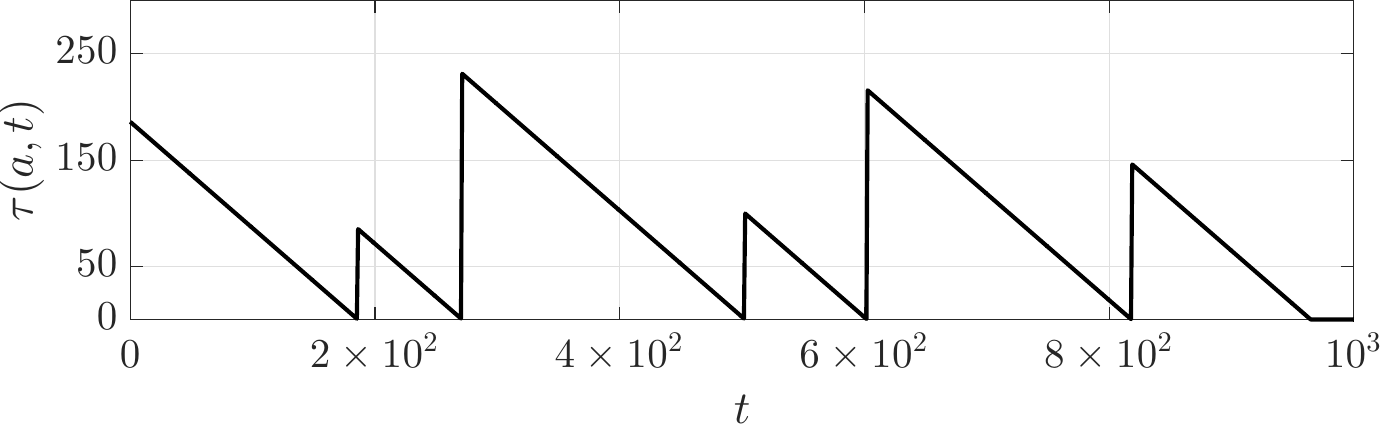}
    \caption{Time evolution of the waiting time $\tau(a,t)$ (see~\eqref{eq:Waiting_Time_Def}) associated to the above timeseries: it is a succession of affine parts with slope $-1$. Note that in principle, there should be small time intervals such that $\tau(a,t)=0$, corresponding to the duration of the event with $\obs(t)>a$, separating the triangles. Here, the duration of the events is too small for such intervals to be visible.}
    \label{fig:illustration_tau}
  \end{subfigure}
  \caption{An example of a random process (a) and the waiting time (b) associated to events reaching a given threshold.}
\end{figure}

We consider a statistically time homogeneous ergodic process (a stationary timeseries) $\left\{ \obs(t)\right\} _{ t\geq t_0}$.
Typically, $\obs:\mathbb{R}^d\to \mathbb{R}$ is an observable on a system of interest, considered here as a $\mathbb{R}^d$-valued stochastic process $\bigl(X_t\bigr)_{t\geq t_0}$, and we should denote $\obs(t)=\obs(X_t)$.
We are interested in the statistical distribution of events where the observable reaches a prescribed threshold $a$.
The occurrence of such events is illustrated for a sample \acf{ou} process, defined by $dX_t=-\alpha X_t dt+\sqrt{2\epsilon}dW_t$, on Fig.~\ref{fig:courbe_signal_avec_seuil}.
We define the return time for a given threshold $a$ as the average time one has to wait before observing the next event with $\obs(t)>a$.
More precisely, we define the waiting time
\begin{align}
  \tau(a,t)&=\min\left\{ \tau\geq t\left|\obs\left(\tau\right)>a\right.\right\} -t. \label{eq:Waiting_Time_Def}\\
  \intertext{As an illustration, the waiting time $\tau(a,t)$ is shown for our sample Ornstein--Uhlenbeck process on Fig.~\ref{fig:illustration_tau}.
  Then the return time $r(a)$ for the threshold $a$ is defined as}
  r(a) &=\mathbb{E}\left[\tau(a,t)\right]\label{eq:Return_Times_Definition},
\end{align}
where $\mathbb{E}$ is the average with respect to realisations of the process $X$ with initial condition $X_{t_0}=x_0$ (hence the notation $\mathbb{E}=\mathbb{E}_{x_0,t_0}$ in that case), or is a time average for an ergodic process.
From now on, we shall omit the indices when there is no ambiguity.
The return time $r(a)$ is independent of time because the process is homogeneous.

The problem we consider in this section is that of estimating $r(a)$ from a sample timeseries of duration $T_{d}: \left\{ \obs(t)\right\} _{0\leq t\leq T_{d}}$.
The definition leads to an obvious estimator for $r(a)$, the \emph{direct estimator} $\hat{r}_D$ defined by
\begin{equation}
  \label{eq:Return_Times_Direct_Estimate} \hat{r}_D(a)=\frac{1}{T_{d}}\int_{0}^{T_{d}}\tau(a,t)\,\text{dt}=\frac{1}{T_{d}}\sum_{n=1}^{N_{d}}\frac{\tau_{n}^{2}}{2},
\end{equation}
where $\tau_{n}$ is the duration of the successive intervals over which $\obs(t)\leq a$, and $N_{d}$ is the number of such intervals.
The last identity in~\eqref{eq:Return_Times_Direct_Estimate} is illustrated graphically in~Fig.~\ref{fig:illustration_tau}: the integral of $\tau(a,t)$ is given by computing the total area beneath the triangles.

In the limit of rare events, the return time will also be the average time between two successive independent events.
However the definition~\eqref{eq:Return_Times_Definition} for the return time has the big advantage of not having to deal with the definition of independent events, which is cumbersome when time correlations are not negligible.
We explain this further in the following section.

\subsubsection{Return times and the distribution of successive events}

Estimating return times using~\eqref{eq:Return_Times_Direct_Estimate} implies computing the time intervals $\tau_n$ between successive events with $\obs(t)>a$.
When $a$ is large enough, most of the times $\obs(t)<a$ and very rarely $\obs(t)>a$.
Then we can distinguish two kinds of contributions to the time intervals $\tau_n$.
On the one hand, we have correlated events corresponding to fluctuations around the threshold value $a$, on a timescale of the order of the correlation time.
From our point of view, these correspond to the same event, with a finite duration.
On the other hand, there are successive events such as those depicted in Fig.~\ref{fig:courbe_signal_avec_seuil}, which can be considered as statistically independent events.
Therefore, we expect those events to form a Poisson point process, and the corresponding time intervals $\tau_n$ should be distributed according to the distribution of time intervals of a Poisson process: $P\left(\tau\right)=\lambda\exp\left(-\lambda\tau\right)$~\cite{EmbrechtsBook,leadbetter_extremes_1983,doucet_sequential_2001}.

\begin{figure}
\centering
\begin{subfigure}{.5\textwidth}
  \centering
  \includegraphics[width=\linewidth]{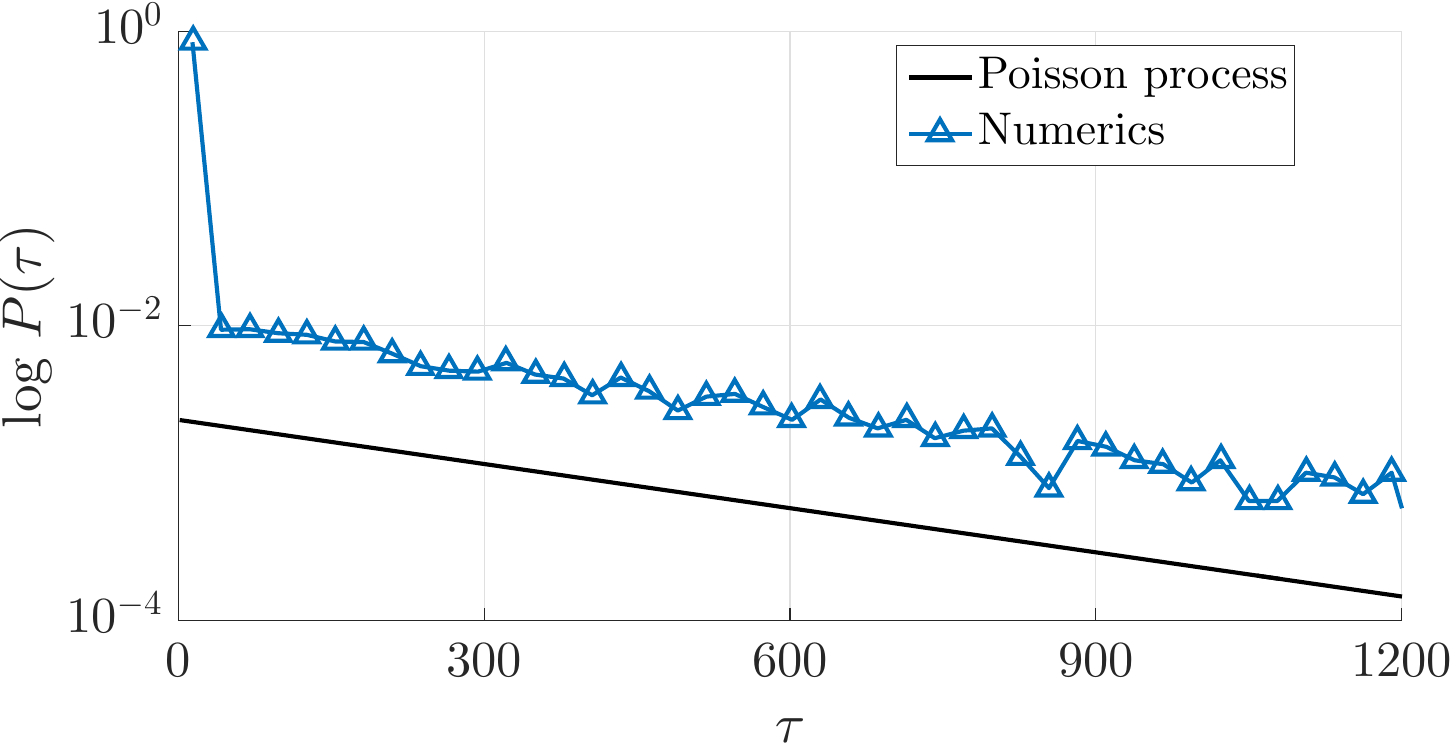}
  \caption{Taking all intervals into account, including those corresponding to oscillations around the threshold.}
  \label{fig:histo_nonpoiss}
\end{subfigure}%
\begin{subfigure}{.5\textwidth}
  \centering
  \includegraphics[width=\linewidth]{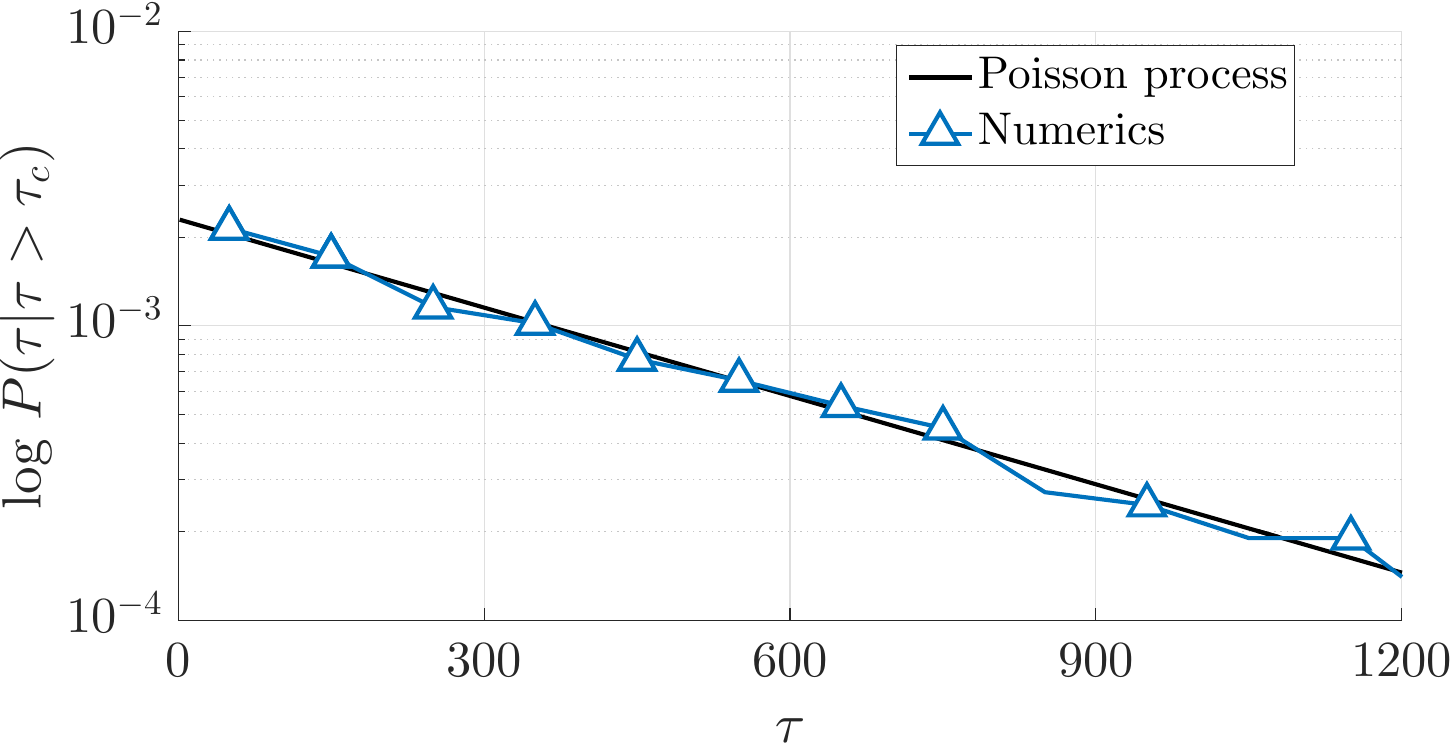}
  \caption{Discarding small intervals ($\tau<\tau_c$) linked to oscillations around the threshold.}
  \label{fig:histo_poiss}
\end{subfigure}
\caption{\ac{pdf} of waiting times between two consecutive fluctuations of amplitude $a = 2.5$, estimated from a timeseries of length $T_d = 10^6$ of the Ornstein--Uhlenbeck process~\eqref{eq:OUprocess} with $\alpha=1$ and $\epsilon=1/2$ (blue triangles), and assuming the events follow a Poisson process with rate $1/r(a)$, $P(\tau) = e^{-\tau/r(a)}/r(a)$ (black solid line), where $r(a)$ is computed from the timeseries.
  The correlation time of the Ornstein--Uhlenbeck process is $\tau_c=1/\alpha=1$.
\label{fig:test}}
\end{figure}
Figure~\ref{fig:histo_nonpoiss} shows the \acf{pdf} of the time interval between two occurrences of an event $\obs(t)>a$, drawn from a sample timeseries generated with an Ornstein--Uhlenbeck process.
One can see that most of the contributions are indeed small intervals of the order of the correlation time.
Discarding all the time intervals below the correlation time, one obtains the \ac{pdf} displayed in Fig.~\ref{fig:histo_poiss}, which coincides with the exponential distribution corresponding to a Poisson point process.

When $a$ is large, $r(a)\gg\tau_{c}$ where $\tau_{c}$ is the correlation time of the process.
Then the contribution of intervals $\tau_n$ of duration comparable to $\tau_{c}$ in the formula~\eqref{eq:Return_Times_Direct_Estimate} becomes asymptotically negligible compared to the contribution of the time intervals $\tau_n \gg \tau_c$.
Graphically, this may be seen as the fact that the sum in~\eqref{eq:Return_Times_Direct_Estimate} is dominated by the contribution of very big triangles, while for small $a$ all the triangles have roughly the same area.
Then, the return time $r(a)$ coincides with the average time between two statistically independent events exceeding the value $a$.
In other words, rare fluctuations can be considered as independent from one another, their duration can be neglected compared to their return time, and the distribution of such events is well approximated by a Poisson process of rate $\lambda = 1/r(a)$.

Neglecting the duration of the extreme events yields $\sum_{n=1}^{N_{d}}\tau_{n}\approx T_{d}$ and then one can check that
\begin{equation} \frac{1}{T_{d}}\sum_{n=1}^{N_{d}}\frac{\tau_{n}^{2}}{2} \approx \frac{N_{d}}{\sum_{n=1}^{N_{d}}\tau_{n}}\frac{1}{N_{d}}\sum_{n=1}^{N_{d}}\frac{\tau_{n}^{2}}{2}\underset{N_{d}\rightarrow\infty}{\rightarrow}\frac{1}{2}\frac{\mathbb{\mathbb{E}}\left\lbrack\tau^{2}\right\rbrack}{\mathbb{\mathbb{E}}\left\lbrack\tau\right\rbrack}=\frac{1}{\lambda(a)}=r(a),
  \label{eq:rate_poisson}
\end{equation}
where the average in this computation is taken with respect to the Poisson process interval \ac{pdf} $P\left(\tau\right)$ made explicit.

One may be tempted to use the estimator $\hat{r}_D'(a)=\frac{1}{N_{d}}\sum_{n=1}^{N_{d}}\tau_{n}$ instead of the estimator $\hat{r}_D$ defined by~\eqref{eq:Return_Times_Direct_Estimate}.
For an actual Poisson process, that would just give the same result.
However this estimator would be more sensitive to the effect of a finite correlation time, since the contributions from time intervals $\tau_n \approx \tau_c$ between successive events will only become negligible linearly in $\tau_c/r(a)$, as opposed to quadratically in formula~\eqref{eq:Return_Times_Direct_Estimate}.

From now on, we shall assume that the statistics of rare events is Poissonian.
This is a reasonable approximation for many dynamical systems as long as there is a well-defined mixing time after which the initial conditions are forgotten.
Of course, it would not hold for systems with long-term memory.
Note that this assumption is similar to the \emph{Independent Interval Approximation} used in the context of persistence~\cite{Bray2013}.
In the next paragraph, we use this assumption to derive new expressions that allow for accurate and efficient sampling of the return times.

\subsubsection{Sampling return times for rare events}
\label{sec:return_time_rare_events}
In this section we present an alternative way to compute return times, that provides an easier and more efficient way to draw return time plots for rare events than using the direct estimator~\eqref{eq:Return_Times_Direct_Estimate}.
Let us divide the timeseries $\left\{ \obs(t)\right\} _{0\leq t\leq T_d}$ in $M$ blocks of duration $\Delta T\gg\tau_{c}$, so that $T_d=M\Delta T$, and let us define the block maximum
\begin{equation}
a_{m}=\max\left\{ \obs(t)\left|(m-1)\Delta T\leq t\leq m\Delta T\right.\right\},
\end{equation}
and $s_{m}(a)=1$ if $a_{m}>a$ and $0$ otherwise, for $1 \leq m \leq M$.

For rare events, \textit{i.e.} $r(a) \gg \tau_c$, the number of events $N(t)=\sum_{m \leq \left\lceil t/\Delta T\right\rceil } s_m(a)$ is well approximated by a Poisson process with density $\lambda(a)=1/r(a)$.
Then, assuming $\tau_{c}\ll\Delta T\ll r(a)$, the probability $q_{m}(a)$ that $a_{m}$ be larger than $a$ is well approximated by $q_{m}(a)\simeq\Delta T/r(a)$.
As $q_{m}(a)$ can be estimated by $\frac{1}{M}\sum_{m=1}^{M}s_{m}(a)$, an estimator of $r(a)$ is the \emph{block maximum estimator}:
\begin{equation}
\hat{r}_B(a) = \frac{T_{d}}{\sum_{m=1}^{M}s_{m}(a)}.\label{eq:Return_Times_Rare}
\end{equation}
This is the classical method for computing the return time of rare events, valid when $\Delta T \ll r(a)$~\cite{Otto2012}.

We now introduce a new, more precise estimator, also valid when $\Delta T/r(a)$ is of order one.
It is obtained by using $q_{m}(a)=1-e^{-\Delta T/r(a)}$.
Then, a better estimator of $r(a)$ is the \emph{modified block maximum estimator}:
\begin{equation}
\hat{r}_B'(a) = -\frac{\Delta T}{\ln\left(1-\frac{1}{M}\sum_{m=1}^{M}s_{m}(a)\right)}.\label{eq:Return_Times_Rare-1}
\end{equation}

\begin{figure}
  \centering
  \includegraphics[width=0.75\linewidth]{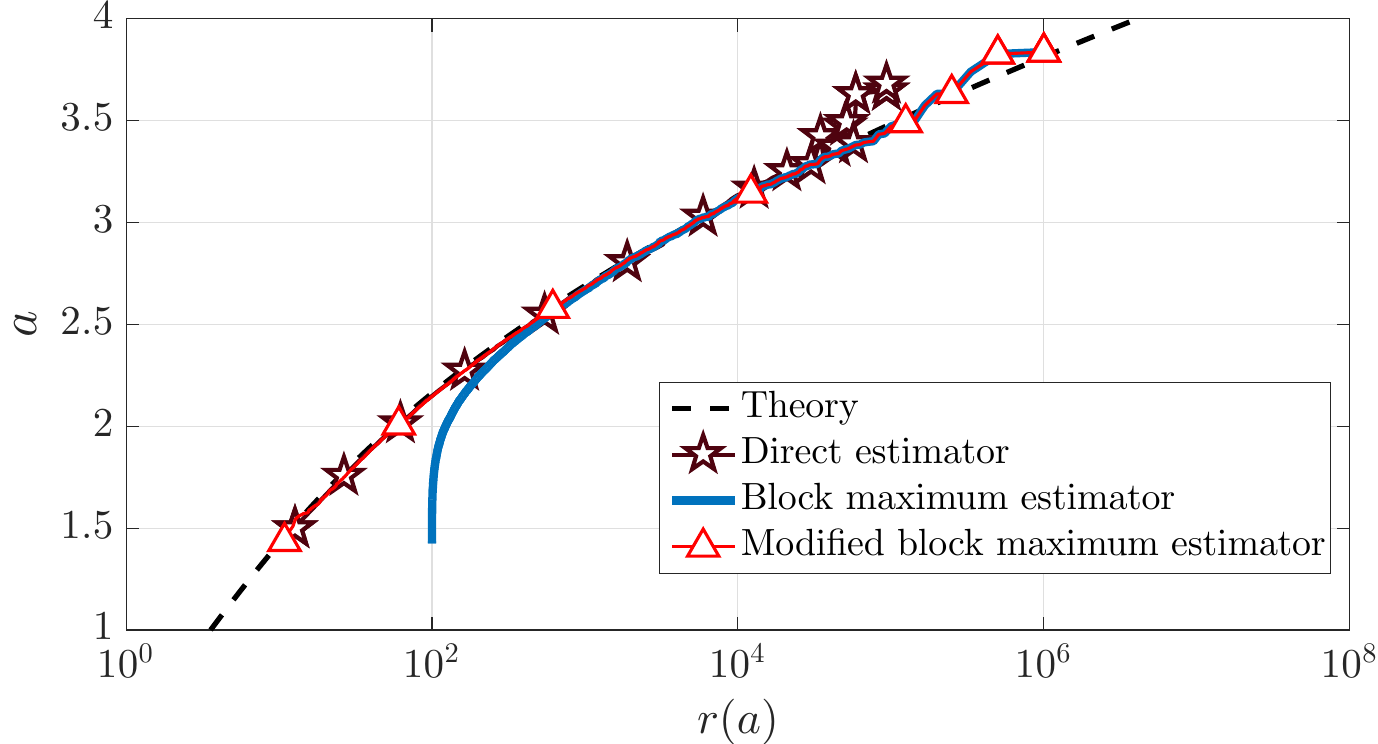}
  \caption{Return time plots for the Ornstein--Uhlenbeck process~\eqref{eq:OUprocess} with $\epsilon = 1/2$, $\alpha = 1$, estimated from a timeseries of length $T_d=10^6$ using the direct estimator $\hat{r}_D$~\eqref{eq:Return_Times_Direct_Estimate} (pentagrams), the block maximum estimator $\hat{r}_B$~\eqref{eq:Return_Times_Rare} ($\Delta T = 100$, solid blue line), and the enhanced block maximum estimator $\hat{r}_B'$~\eqref{eq:Return_Times_Rare-1} ($\Delta T = 100$, solid red line and white triangles).
    These estimates are compared to the analytical solution~\eqref{eq:MeanFirstPassage_S} (dashed black line).}
  \label{fig:temps_retour_OU_direct}
\end{figure}

To compute these estimators in practice, we sort the sequence $\left\{ a_{m}\right\} _{1\leq m\leq M}$ in decreasing order and denote the sorted sequence $\left\{ \tilde{a}_{m}\right\} _{1\leq m\leq M}$ such that $\tilde{a}_{1}\geq\tilde{a}_{2}\geq...\geq\tilde{a}_{M}$.
Based on \eqref{eq:Return_Times_Rare}, we then associate to the threshold $\tilde{a}_{m}$ the return time $r(\tilde{a}_{m})=M\Delta T/m$.
Indeed, $\sum_{\ell=1}^{M}s_\ell(\tilde{a}_m)=m$, which means that $m$ events with amplitude larger than $\tilde{a}_{m}$ have been observed over a duration $M\Delta T$.
Alternatively, using the more precise estimator $\hat{r}_B'$~\eqref{eq:Return_Times_Rare-1} we associate to the threshold $\tilde{a}_{m}$ the return time $r\left(\tilde{a}_{m}\right)=-\frac{\Delta T}{\log\left(1-\frac{m}{M}\right)}$.
The return time plot represents $\tilde{a}_{m}$ as a function of $r\left(\tilde{a}_{m}\right)$, as illustrated for instance on Fig.~\ref{fig:temps_retour_OU_direct}.
Let us stress again that formulas~\eqref{eq:Return_Times_Rare} and~\eqref{eq:Return_Times_Rare-1} and this method of plotting the return time are meaningful only if doing block maxima, and for ranges of parameters such that $\tau_{c}\ll\Delta T\ll r(a)$ for \eqref{eq:Return_Times_Rare} or $\tau_c \ll \Delta T$ for~\eqref{eq:Return_Times_Rare-1}.

Figure~\ref{fig:temps_retour_OU_direct} illustrates the three methods for computing return times from a timeseries: from the definition~\eqref{eq:Return_Times_Direct_Estimate} and the two formulas~\eqref{eq:Return_Times_Rare} and~\eqref{eq:Return_Times_Rare-1}.
The sample timeseries used in this figure is extracted from an Ornstein--Uhlenbeck process, for which the return time curve can also be computed analytically.
One can see that both formulas~\eqref{eq:Return_Times_Rare} and~\eqref{eq:Return_Times_Rare-1} lead to the same estimate for events with $r(a) \gg \Delta T$.
However, formula~\eqref{eq:Return_Times_Rare} fails to yield a correct estimate as soon as $r(a) \simeq \Delta T$.

For rare events, plotting return times using~\eqref{eq:Return_Times_Rare}, as is classically done, proves itself much more convenient and efficient than the naive sampling using~\eqref{eq:Return_Times_Direct_Estimate}.
It is important to note however, that the use of~\eqref{eq:Return_Times_Rare} is valid only after computing maxima over an interval of duration $\Delta T$ much larger than $\tau_{c}$, a remark that not been considered in many previous publications.
Moreover, the generalisation~\eqref{eq:Return_Times_Rare-1} we propose in this paper is much more accurate for events with a return time of order of $\Delta T$.
This procedure to compute return time plots can also be generalised in combination with the use of rare event algorithms, as we shall see in the next section.

\subsection{Computing return times from a rare event algorithm}
\label{sec:return_time_algorithm}

In section~\ref{sec:return_time_timeseries}, we defined the return time for a time-homogeneous stochastic process and explained how to efficiently compute it for rare events from a timeseries.
However, a major difficulty remains as we still have to generate numerically the rare events in the timeseries, which comes at a large computational cost.
In the present section, we explain how to apply the above method to the data produced by algorithms designed to sample efficiently rare events instead of direct simulations.

Rare event algorithms provide an effective ensemble of $M$ trajectories $\{X_{m}(t)\}_{0\leq t \leq T_a}$ ($1 \leq m \leq M$).
Note that the length $T_a$ of the trajectories generated by the algorithm does not necessarily coincide with the length $T_d$ of the trajectory generated by direct sampling: in practice, as we shall see, $T_a \ll T_d$.
For each of these trajectories, we compute the maximum of the observable over the time evolution $a_{m}=\max_{0\leq t\leq T_{a}}\left( A(X_{m}(t))\right)$.
This is similar to the block maximum method described in section~\ref{sec:return_time_rare_events}, with each trajectory playing the role of a block.
There is however a major difference: unlike in the block maximum method, the different trajectories sampled by the rare event algorithm do not have identical statistical weight.
To each trajectory $X_m(t)$, and thus to each maximum $a_m$, is associated a probability $p_{m}$ computed by the algorithm.
Hence, rather than just a sequence $\left\{ a_{m}\right\} _{1\leq m\leq M}$, rare event algorithms yield a sequence $\left\{ a_{m},p_{m}\right\} _{1\leq m\leq M}$.
The generalisation of the block maximum formula~\eqref{eq:Return_Times_Rare-1} to non-equiprobable blocks is straightforward and leads to the estimator
\begin{equation}
\hat{r}_A(a)=-\frac{T_{a}}{\ln\left(1-\sum_{m=1}^{M}p_{m}s_{m}(a)\right)}.\label{eq:Return_Time_Large_Deviation_Algorithm-1}
\end{equation}
Of course, we could construct similarly an estimator generalising~\eqref{eq:Return_Times_Rare}, but as we have seen in the previous section, the estimator~\eqref{eq:Return_Times_Rare-1} yields better performance.

In practice, to plot the return time curve, we sort the sequence $\left\{ a_{m},p_{m}\right\} _{1\leq m\leq M}$ in decreasing order with respect to the $a_m$, and denote the sorted sequence $\left\{ \tilde{a}_{m},\tilde{p}_{m}\right\} _{1\leq m\leq M}$ such that $\tilde{a}_{1}\geq\tilde{a}_{2}\geq...\geq\tilde{a}_{M}$.
We then associate to the threshold $\tilde{a}_{m}$ the return time
\begin{equation}
\hat{r}_A(\tilde{a}_{m})=-\frac{T_{a}}{\ln\left(1-\sum_{\ell=1}^{m}\tilde{p}_{\ell}\right)}.\label{eq:Return_Time_Large_Deviation_Algorithm}
\end{equation}
Indeed, the sum of the weights of the events with amplitude larger than $\tilde{a}_m$ is $\sum_{\ell=1}^{m}\tilde{p}_\ell$.
Again, the return time plot represents $a$ as a function of $r\left(a\right)$.

We stress that the method described here does not depend on the observable of interest, or on the details of the algorithm itself.
In the remainder of the paper, we provide a \emph{proof-of-concept} for this method, by considering two kinds of observables, sampled by two different algorithms: first, we study the return times for instantaneous observables using the \acf{ams} algorithm (section~\ref{sec:AMS}), then we turn to time-averaged observables using both the \ac{ams} and the \acf{gktl} algorithm (section~\ref{sec:gktl}).
We show that the method allows to accurately compute return times at a much smaller computational cost than direct simulation.
In both cases, we apply the technique to the simple case of an Ornstein--Uhlenbeck process, for which the results are easily compared with direct simulation and theoretical predictions, before illustrating the potential of the method for applications in complex systems (section~\ref{sec:applications}).

\section{Return times sampled with the Adaptive Multilevel Splitting algorithm}
\label{sec:AMS}

In this section, we present the computation of return times by applying the method presented in section~\ref{sec:return_time_algorithm} to a rare event algorithm known as \acf{ams}.
This algorithm follows the strategy of \emph{splitting methods} for the estimation of rare event probabilities, which dates back to the 1950s~\cite{KahnHarris1951}.
Many variants have been proposed since then.
The \ac{ams} algorithm can be interpreted as simulating a system $\{x_i(t)\}$ of interacting replicas (instead of independent replicas in a crude Monte Carlo simulation), with some \emph{selection and mutation} mechanism.
We describe this mechanism in section~\ref{sec:amsalgo_rettimes} as a method to sample trajectory space.
This contains all the necessary details for practical use of the algorithm.
Then, in section~\ref{sec:amsalgo} we connect the procedure to the general framework of the \ac{ams} algorithm, which enables us to directly benefit from the available mathematical results.
In section~\ref{sec:committor}, we explain what is the optimal choice of score function for our problem and we analyse its behaviour.
In section~\ref{sec:ams_returntimes}, we show how the algorithm enables us to estimate return times, under the Poisson statistics assumption made above.
Finally, we illustrate in section~\ref{sec:return_time_AMS} the method by computing the return times for an Ornstein--Uhlenbeck process.

\subsection{The \acf{tams} algorithm}
\label{sec:amsalgo_rettimes}

The classical \ac{ams} algorithm is based on the evolution of an ensemble of trajectories, based on selection-mutation rules, in order to compute rare event probabilities, and more generally committor functions.
Return times can not be estimated directly from a committor function and require the estimation of trajectory statistics.
The method we propose to compute return times involves the estimation of probabilities of trajectories with a fixed duration $T_a$.
In order to deal with this, we propose a specific modification of the classical \ac{ams} algorithm, called Trajectory Adaptive Multilevel Splitting.

While the classical \ac{ams} algorithm requires to specify only a real-valued \emph{score function} $\xi$ -- also called a \emph{reaction coordinate} in many works, due to connections with molecular dynamics simulations, see~\cite{Cerou2011}, and also~\cite[Section~4.3]{Brehier2016a} -- the Trajectory Adaptive Multilevel Splitting requires in general a time dependent score function, see Section~\ref{sec:committor} for the optimal choice.

We consider a continuous time Markov model able to generate trajectories.
It can be either a stochastic process, for instance a diffusion, or a chaotic deterministic dynamical system.
Let us now describe the algorithmic procedure.

We start by simulating $\nsample$ independent trajectories, denoted $\{x_n^{(0)}(t)\}_{1\leq n \leq \nsample}$, for a fixed duration $T_a$.
To each of these trajectories, we associate a weight $w_0=1$.
Then, at iteration $j\geq 1$, we evaluate the performance of all replicas $\{x_n^{(j-1)}(t)\}_{1 \leq n \leq \nsample}$ at iteration $j-1$, measured by the maximum of the score function $\xi$ over the whole trajectory:
\begin{equation}
  \mathcal{Q}_n^{(j)} = \sup_{0 \leq t \leq T_a} \xi(t,x_n^{(j-1)}(t)).
\end{equation}
We select the trajectories corresponding to the lowest $\mathcal{Q}_n^{(j)}$: let us denote $\mathcal{Q}_j^\star= \min_{1\leq n \leq \nsample} \mathcal{Q}_n^{(j)}$ and $n_{j,1}^\star,\ldots,n_{j,\ell_j}^\star$ the indices such that:
\begin{equation}
  \mathcal{Q}_{n_{j,1}^\star}^{(j)} = \cdots = \mathcal{Q}_{n_{j,\ell_j}^\star}^{(j)} = \mathcal{Q}_j^\star.
\end{equation}
One might expect intuitively that $\ell_j=1$.
This is not necessarily the case, as explained in~\cite{Brehier2016a}: because of the discretization of the dynamical equations in the numerical model, two or more trajectories may yield the same level $\mathcal{Q}_n^{(j)}$.

\begin{figure}[!h]
  \centering
  \includegraphics[scale=0.50]{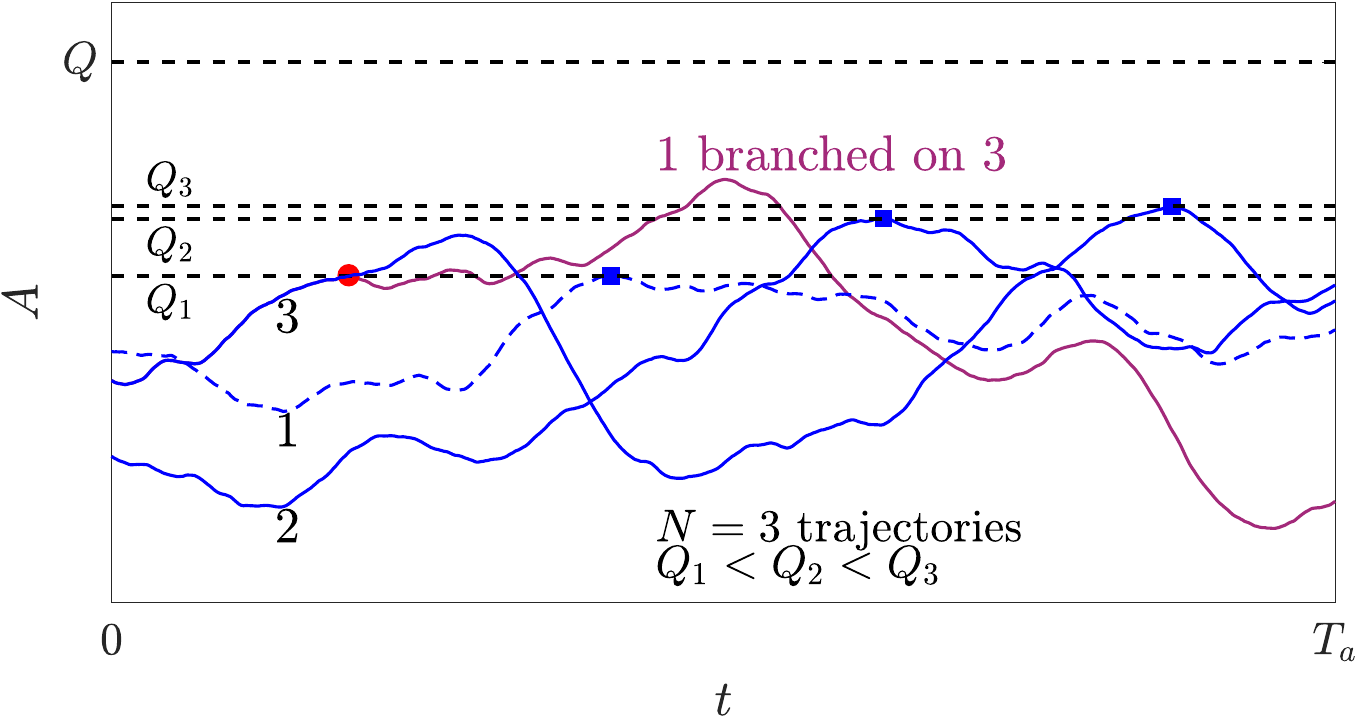}
    \caption{Illustration of one selection-mutation step in the \ac{ams} algorithm for the computation of the probability that an observable $\obs:\mathbb{R}^d\to \mathbb{R}$  reaches values larger than $Q$ over a trajectory of duration $T_a$.}
  \label{fig:AMS_schema}
\end{figure}

We then proceed to the mutation step.
For each trajectory $x_{n_{j,\ell}^\star}^{(j-1)}$ ($1 \leq \ell \leq \ell_j$), we choose a trajectory $x_{n_\ell}^{(j-1)}$ ($n_{\ell} \neq n_{j,1},\ldots n_{j,\ell_j}$) randomly among the $\nsample-\ell_j$ remaining trajectories, and define the time $t_{j,\ell}$ defined as the smallest time $t$ such that $\xi(t,x_{n_\ell}^{(j-1)}(t))>\mathcal{Q}_j^\star$.
Finally, we define the new replica $x_{n_{j,\ell}^\star}^{(j)}$ by copying the trajectory $x_{n_\ell}^{(j-1)}$ from $t_0$ to $t_{j,\ell}$, and simulating the rest of the trajectory, from $t_{j,\ell}$ to $T_a$.
For a Markov process, for instance a diffusion, a new realisation of the noise is used in order to simulate the new trajectory from $t_j$ to $T_a$.
For a chaotic deterministic system, a small amplitude noise is added to the initial condition at time $t_j$.
The other trajectories are not modified: $x_n^{(j)}=x_n^{(j-1)}$ for $n \neq n_{j,1}^\star,\ldots,n_{j,\ell}^\star$.
The selection-mutation process is illustrated on Fig.~\ref{fig:AMS_schema}.
We associate to the trajectories $x_n^{(j)}$ forming the ensemble at step $j$ the weight $w_j$ given by~\cite{Cerou2007,Cerou2011,Brehier2016a}:
\begin{equation}
  w_j = \prod_{i=1}^j \left( 1 - \frac{\ell_i}{\nsample}\right)=\left( 1 - \frac{\ell_j}{\nsample}\right)w_{j-1}.
\end{equation}
Note that we could mutate more replicas at each step by selecting an arbitrary number of levels $\mathcal{Q}_n^{(j)}$, instead of just the minimum $\mathcal{Q}_j^\star$ as described above.
The particular case described above is sometimes referred to as the \emph{last particle method}~\cite{Simonnet2016}.

The selection-mutation process is iterated $\niter$ times (two possible definitions of $\niter$ are given below).
The number of resampled trajectories is given by $\tilde{\niter} = \sum_{j=1}^\niter \ell_j$.
Note that $\tilde{\niter} \geq \niter$, but the two need not necessarily coincide.
In the end, the algorithm generates $M=\nsample+\tilde{\niter}$ trajectories, given explicitly by the set $\{x_n^{(0)}\}_{1 \leq n \leq \nsample} \cup \{ x_{n_{j,\ell}^\star}^{(j)}\}_{1 \leq \ell \leq \ell_j, 1 \leq j \leq \niter}$, or equivalently, the set $\{x_n^{(\niter)}\}_{1 \leq n \leq \nsample} \cup \{ x_{n_{j,\ell}^\star}^{(j-1)}\}_{1 \leq \ell \leq \ell_j, 1 \leq j \leq \niter}$.
Each trajectory has an associated weight, given by the iteration until which it was a member of the ensemble: $w_J$ for the final trajectories $\{x_n^{(\niter)}\}_{1 \leq n \leq \nsample}$, and $w_{j-1}$ for the trajectories $\{ x_{n_{j,\ell}^\star}^{(j-1)}\}_{1 \leq \ell \leq \ell_j, 1 \leq j \leq \niter}$ mutated at iteration $1 \leq j \leq \niter$.
Let us relabel these trajectories and their associated weights as $\{ (x_m,w_m)\}_{1 \leq m \leq M}$.
Normalising the weights with $W=\sum_{m=1}^M w_m$, we obtain the probabilities $p_m=w_m/W$ associated with the trajectories.

Note that instead of just one realisation of the algorithm, one may carry out $\nexp$ independent realisations, thus yielding $M=\sum_{k=1}^\nexp (\nsample_k+\tilde{\niter}_k)$ trajectories with the associated weights, where $\nsample_k$ and $\tilde{\niter}_k$ denote the number of initial trajectories and resampled trajectories for realisation $k$, respectively.
The probabilities for the trajectories are computed as above.

For any observable $O[x(t)]$, we can define an estimator based on our sampling of trajectory space:
\begin{equation}
  \hat{O}_M = \sum_{m=1}^M p_m O[x_m(t)].\label{eq:amsestimator}
\end{equation}

For practical applications, we shall be interested in two particular cases:
\begin{itemize}
\item Instantaneous observable: $O[X,t]=\obs(X(t))$, for some time-independent observable $\obs: \mathbb{R}^d \to \mathbb{R}$.

\item Time-averaged observable: $O[X,t]=\frac{1}{T}\int_{t-T}^{t}\obs(X(s))ds$ for some time-independent observable $\obs: \mathbb{R}^d \to \mathbb{R}$ and prescribed width $T$ for the averaging window.
  Note that this is a case where the time-dependent observable $O$ is defined on a different interval than the original process $X$, here $[T,T_a]$.

\end{itemize}

The number of iterations $\niter$ can either be a prescribed integer (in that case the stopping criterion for the algorithm is simply $j=\niter$), or a random number such that all the trajectories in the ensemble reach a threshold level $\mathcal{Q}$ (the stopping criterion is then $\mathcal{Q}_n^{(j)}>\mathcal{Q}$ for all $1 \leq n \leq \nsample$).
The latter case is more common in existing \ac{ams} implementations, however both cases are covered by the general framework developed in~\cite{Brehier2016a}, and give consistent results.
We further discuss these two possible choices in section~\ref{sec:ams_returntimes}.

Let us now estimate the computational cost of an \ac{ams} run.
The number of trajectories generated by an \ac{ams} run is $M=\nsample+\tilde{\niter}$, as pointed out above.
Each resampled trajectory is not simulated over the whole duration $T_a$, but over $\tau<T_a$, with $\tau$ a random number depending on the branching point.
We thus define $\gamma \in [0,1]$ so that $\mathbb{E}[\tau]=\gamma T_a$ is the average duration of the resampled part of a mutated trajectory.
Performing $\nexp$ identical and independent realisations of the \ac{ams} algorithm, the average computational cost associated with a given experiment is then approximately
\begin{equation}
  \label{eq:computationalcost_ams}
  \mathcal{C} = \nexp \times (\nsample + \gamma \niter)T_a.
\end{equation}

\subsection{Connection with the Adaptive Multilevel Splitting (AMS) algorithm for time-dependent observables}
\label{sec:amsalgo}

In this section, we describe the connection between the \acf{tams} algorithm and the classical \ac{ams} algorithm.
The aim is to deduce the mathematical properties of the \ac{tams} algorithm from the known ones for the \ac{ams} algorithm.
For instance we will conclude that the optimal score function is the committor function (\ref{eq:time_dependent_committor}).
This section can be skipped by the reader interested in the algorithm only, without trying to understand the mathematical aspects.

The \acf{ams} algorithm has originally been designed~\cite{Cerou2007} to efficiently and accurately estimate probabilities of rare events of the type $\mathbb{P}_{x_0,t_0}(\tau_{\set{B}}<\tau_{\set{A}})\in(0,1)$: the probability that a Markov process $\bigl(X_t\bigr)_{t\ge t_0}$, initialised with $X_{t_0}=x_0$, hits a set $\set{B}$ before hitting a set $\set{A}$ (with $\set{A}\cap \set{B}=\emptyset$), where $\tau_{\set{C}}=\inf\left\{t> t_0; X_t\in \set{C}\right\}$ is the hitting time of a set $\set{C}$.
In this section, we show how the problem of estimating the maximum value of a time-dependent observable over a trajectory (which later will be used to estimate return times) falls within the scope of the \ac{ams} algorithm.
This enables us to benefit directly from the theoretical properties of the \ac{ams} algorithm.
Some recent mathematical results about the algorithm are reviewed in appendix~\ref{sec:AMSproperties}.
This review is not exhaustive, see for instance~\cite{Brehier2016a} and references therein.

We consider a $\mathbb{R}^d$-valued Markov process $\bigl(X_t\bigr)_{t\in[0,T_a]}$, with continuous trajectories, for some fixed final time $T_a$, and a time-dependent observable $O[X,t]$: this is a (time-dependent) functional of the process $X$, taking value in $\mathbb{R}$.
It may be defined for times belonging to a subset of $[0,T_a]$, but for simplicity we shall still denote $T_a$ the final time.
The aim is to estimate the probability that the observable reaches a threshold $a$ at some point of the trajectory, i.e.
\begin{equation}
  q(a) = \mathbb{P}_{x_0,0} \left \lbrack \underset{0\le t\le T_a}\max O[X,t] > a \right\rbrack;
\end{equation}
(the notation $\mathbb{P}_{x_0,t_0}$ means the probability over realisations of the Markov process with initial condition $X_{t_0}=x_0$).
The \ac{ams} algorithm provides an estimator $\hat{q}(a)$ for this quantity.
Indeed, the event $\left\{\max_{0\le t\le T_a} O[X,t] > a\right\}$ can be identified with the event $\left\{\tau_\set{B}<\tau_\set{A}\right\}$ for an auxiliary Markov process $Y_t$, with an appropriate definition of the sets $\set{A}$ and $\set{B}$, as follows:
\begin{equation}
  Y_t=(t,O[X,t])\in[0,T_a]\times \mathbb{R},\qquad \set{A}=\left\{(T_a,z);~z\le a\right\},\qquad \set{B}=\left\{(t,z);~t\in[0,T_a], z>a\right\}.
\end{equation}
Note that $Y$ is not necessarily a time-homogeneous process.
In section~\ref{sec:amsalgo_rettimes}, we have described the \ac{tams} algorithm that gives a procedure to sample the process $Y$ to provide a good estimate of $q(a)$, based on a score function $\xi$, which measures the distance between $\mathcal{A}$ and $\mathcal{B}$ (in many implementations of the \ac{ams}, $\xi(\partial \mathcal{A})=0$ and $\xi(\partial\mathcal{B})=1$).
We describe the corresponding estimator $\hat{q}(a)$, and the related estimator for return times, in section~\ref{sec:ams_returntimes}.

It follows from the above paragraph that the convergence properties of the \ac{tams} algorithm are a direct consequence of the known results for the \ac{ams} algorithm (see appendix~\ref{sec:AMSproperties}).
Let us, however, explain in a heuristic way the validity of the algorithm.
We refer to~\cite{Cerou2007,Brehier2016a} for rigorous mathematical arguments.

The algorithm iterates a selection-mutation mechanism on a system of clones.
At the selection step, (typically) one clone is removed from the system.
To keep a constant number of clones, one new replica needs to be sampled.
Statistical consistency is ensured by the introduction of the weights, and appropriate rules for their update.
Observe in particular that the sum of the weights of the $\nsample-1$ selected clones, before update, is equal to the sum, after update, of the $\nsample$ clones.
At the mutation step, the new clone is sampled by branching one of the selected clones, at the current level.
The Markov property of the dynamics is used to sample the end of the trajectory, after crossing the current level.
This ensures that, after the mutation step, the $\nsample$ clones observed after the first crossing of the current level, are (conditionally) independent and identically distributed.
Observe that they also have the same weight.
Eventually, at the last iteration, all the $\nsample$ clones reach the (rare) event of interest, by construction.
The weights (or equivalently the random number of iterations) are used to estimate the probability of this event.
First, consider non-adaptive versions of the multilevel splitting algorithm, where the levels and the number of iterations are fixed, as originally developed in~\cite{KahnHarris1951}.
This consists in a decomposition of the probability of the rare event, as a product of conditional probabilities.
The weights are then products of standard estimators of these conditional probabilities.
In the adaptive versions, initially developed in~\cite{Cerou2007}, the levels are computed on-the-fly as empirical quantiles: the minima of scores among $\nsample$ clones.
The factor $1-1/\nsample$ can be interpreted as the associated conditional probability, hence the validity of the approach -- but the analysis in the adaptive case is more complex.

\subsection{The optimal score function}
\label{sec:committor}
This section is a theoretical discussion of the properties of the optimal score function; it may be skipped by readers who are only interested in the application of the \ac{tams} algorithm for computing return times.

%
%
As explained in appendix~\ref{sec:AMSproperties}, the statistical properties, and in particular the variance of the \ac{ams} estimator $\hat{q}(a)$, depend on the choice of the score function $\xi$.
The variance is minimal for a particular choice of the score function, sometimes referred to as the \emph{committor}.
In a very generic manner, for the \ac{ams} algorithm, it is given by $\bar{\xi}=\mathbb{P}[\tau_\set{B}<\tau_\set{A}]$.
In the specific case of the \ac{tams} algorithm, the optimal score function takes the form:
\begin{equation}
  \bar{\xi}(t,x;T_a,a)=\mathbb{P}_{x,t}\left\lbrack\underset{t\le s\le T_a}\max O[X,s] > a\right\rbrack,
  \label{eq:time_dependent_committor}
\end{equation}
for all $(t,x)\in[0,T_a]\times \mathbb{R}^d$, where we denote $\mathbb{P}_{x,t}$ the probability over the process initialised at position $x$ at time $t$, and the threshold $a$ and trajectory duration $T_a$ are fixed parameters.
Note that the optimal score function depends both on time and space.
Of course, we cannot use this score function in practice, because it is exactly what we are trying to compute.
Indeed, as mentioned above, the algorithm ultimately provides an estimate of the probability $q(a) =\bar{\xi}(0,x_0;T_a,a)$.
Nevertheless, a crucial point to implement the \ac{ams} algorithm is to choose a score function that provides a good approximation of the committor.
In practical applications, constructing the score function will often be based on heuristic considerations, but it may also be useful to have theoretical results about the optimal score function.

Here, we want to explain the qualitative properties of the time-dependent committor~\eqref{eq:time_dependent_committor} specific to the \ac{tams} algorithm.
For simplicity, we shall only discuss the case of an instantaneous observable: $O[X,t]=\obs(X_t)$.
Moreover, for the precision of the discussion, we assume that the stochastic process $X$ solves the stochastic differential equation $dX_t = b(X_t)dt+\sqrt{2\epsilon}dW_t$, where $b$ is a vector field with a single fixed-point $x_\star$.
We further assume that the basin of attraction of $x_\star$ is the full phase space.
With this hypothesis, the invariant measure of the diffusion is concentrated close to the attractor $x_\star$ when $\epsilon \ll 1$.
Let us assume that the set $\mathcal{C} = \{ x \mid \obs(x) \leq 0\}$ is a neighbourhood of $x_\star$ on which most of the invariant measure mass is concentrated.
We call $\mathcal{C}$ the attractor.
The target set $\mathcal{D} = \{ x \mid \obs(x) \geq a \}$ is similarly defined.
The hitting times for the sets $\mathcal{C}$ and $\mathcal{D}$ are the random variables given by $\tau_\star = \inf \{ t>0 \mid \obs(X_t) \leq 0 \}$ and $\tau_a = \inf \{ t>0 \mid \obs(X_t) \geq a \}$, respectively, where the process is started from a point $x$ at time $t=0$, such that $0\leq A(x)\leq a$.
We finally define the static committor $\xi_0(x,a) \equiv \mathbb{P}_{x,0}[\tau_a < \tau_\star]$.
The aim of the following discussion is to explain the relation between the time-dependent committor~\eqref{eq:time_dependent_committor} and the static committor $\xi_0(x,a)$.

On the one hand, the time-dependent committor $\bar{\xi}$ satisfies a backward Fokker-Planck equation
\begin{equation}
  \frac{\partial \bar{\xi}}{\partial t} = - L[\bar{\xi}], \quad \text{ with } L=b_i\frac{\partial}{\partial x_i}+\epsilon \frac{\partial^2}{\partial x_i^2},\label{eq:committor-fp}
\end{equation}
in the domain $\obs^{-1}([0,a]) \subset \mathbb{R}^d$ with boundary condition $\bar{\xi}(t,x;T_a,a)=1$ for $x \in \partial \mathcal{D}$, and final condition $\bar{\xi}(T_a,x;T_a,a)=0$.
This follows directly from the backward Fokker-Planck equation for the transition probability $P(y,s|x,t)$, and the fact that, with an absorbing boundary condition on $\partial \mathcal{D}$, $\bar{\xi}(t,x;T_a,a)=1-\int dy P(y,T_a|x,t)$.
Note that when $T_a-t \gg r(a)$, $\bar{\xi}(t,x;T_a,a)\approx 1$ everywhere ($\bar{\xi}$ converges to $1$).
On the other hand, $\xi_0(x,a)$ satisfies $L[\xi_0]=0$, but with different boundary conditions: $\xi_0(x,a)=1$ if $x \in \partial \mathcal{D}$ and $\xi_0(x,a)=0$ if $x\in \partial \mathcal{C}$.
In the next paragraph, we argue that when $T_a-t$ is much smaller than $r(a)$, the time-dependent committor $\bar{\xi}(t,x;T_a,a)$ given by~\eqref{eq:time_dependent_committor} is well approximated by the static committor $\xi_0(x,a)$, except in two boundary layers: a spatial one of size $\epsilon$ for $x$ close to the attractor, and a temporal one of size $\tau_c$ for $t$ close to $T_a$.

Using the notations of section~\ref{sec:amsalgo}, the events $\{ \tau_\set{B} < \tau_\set{A}\}$ can be decomposed into the disjoint union of events for which the observable reaches the threshold $a$ before or after hitting $0$.
The typical time for $X$ to reach $\mathcal{C}$ is the correlation time $\tau_c$.
If we assume that $T_a-t \gg \tau_c$, we have the approximation $\bar{\xi}(t,x;T_a,a) \simeq \xi_0(x,a) + \lbrack 1-\xi_0(x,a) \rbrack \bar{\xi}(t,x_\star;T_a,a)$ (we have used here the approximations  $\bar{\xi}(\tau_\star,y;T_a,a) \simeq \bar{\xi}(\tau_\star,x_\star;T_a,a)$ for any $y\in \partial \mathcal{C}$, and $\bar{\xi}(\tau_\star,x_\star;T_a,a) \simeq \bar{\xi}(t,x_\star;T_a,a)$).
Moreover, when $T_a-t \ll r(a)$, the Poisson approximation $\bar{\xi}(t,x_\star;T_a,a)\simeq (T_a-t)/r(a)$ holds.
To sum up, in the limit $\tau_c \ll T_a-t \ll r(a)$,
\begin{equation}
  \bar{\xi}(t,x;T_a,a) \simeq \xi_0(x,a) + \frac{T_a-t}{r(a)}\lbrack 1-\xi_0(x,a) \rbrack.
\end{equation}
Let us now introduce the quasipotential $V$.
We note that $\xi_0(x,a) \underset{\epsilon \rightarrow 0}{\asymp} \exp(-(\inf_{y \in A^{-1}(\{a\})} V(y)-V(x))/\epsilon)$, while $r(a) \underset{\epsilon \rightarrow 0}{\asymp} \exp((\inf_{y \in A^{-1}(\{a\})} V(y))/\epsilon)$.
We can thus conclude that $\xi_0(x,a)$ dominates this expression for all $x$ except in a region of size $\epsilon$ around the attractor $x_\star$.

\begin{figure}[ht]
  \centering
  \includegraphics[width=0.5\linewidth]{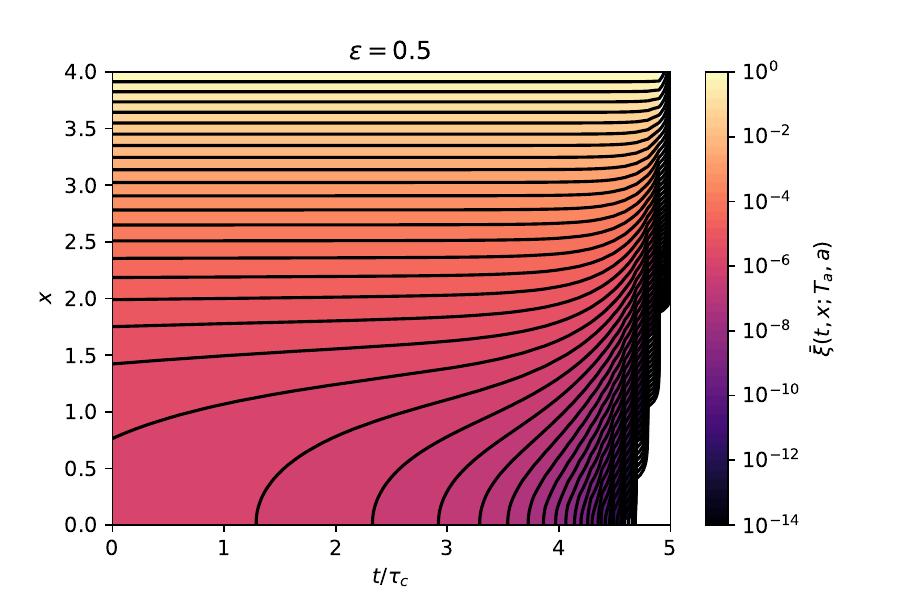}
  \caption{\label{fig:committor} Contour lines of the time-dependent committor $\bar{\xi}(t,x;T_a,a)$ for the Ornstein--Uhlenbeck process (with $\alpha=1,\epsilon=1/2$; in particular $\tau_c=1$), obtained by solving numerically the backward Fokker-Planck equation~\eqref{eq:committor-fp}, with $a=4,T_a=5$.}
\end{figure}
As a conclusion, when $T_a-t$ is much smaller than $r(a)$, the time-dependent committor $\bar{\xi}(t,x;T_a,a)$~\eqref{eq:time_dependent_committor} is well approximated by the static committor $\xi_0(x,a)$, except in two boundary layers: a spatial one of size $\epsilon$ for $x$ close to the attractor, and a temporal one of size $\tau_c$ for $t$ close to $T_a$.
This is illustrated in Fig.~\ref{fig:committor}, representing the committor $\bar{\xi}(t,x;T_a,a)$ for the Ornstein--Uhlenbeck process (with $\alpha=1,\epsilon=1/2$), obtained by solving numerically the backward Fokker-Planck equation~\eqref{eq:committor-fp}, with $a=4,T_a=5$.

\subsection{Computing return times}\label{sec:ams_returntimes}
%
%
As explained in section~\ref{sec:amsalgo_rettimes}, the algorithm generates an ensemble of $M$ trajectories $x_m(t)$ with associated probability $p_m$.
It follows directly from~\eqref{eq:amsestimator} that an estimator of $q(a)$ is:
\begin{equation}
  \hat{q}_M(a) = \sum_{m=1}^M p_m s_m(a),\label{eq:ams_estimator_q}
\end{equation}
where $a_m=\max_{0\leq t \leq T_a} \obs(x_m(t))$ is the maximum value for the observable over the trajectory $m$, $p_m$ the associated probability (see~\ref{sec:amsalgo_rettimes}), and $s_m(a)=1$ if $a_m>a$, $0$ otherwise (\ref{sec:return_time_rare_events}).

As explained in~\ref{sec:return_time_rare_events}, the return time is related to $q(a)$ by the hypothesis that these events are Poissonian, and we obtain the estimator for the return time $\hat{r}_M(a)=\frac{T_a}{\ln(1-\hat{q}_M(a))}$ given by~\eqref{eq:Return_Time_Large_Deviation_Algorithm-1} (alternatively, we could use $\hat{r}_M(a)=\frac{T_a}{\hat{q}_M(a)}$).
In essence, to draw return time plots, it suffices to sort the set $\left\{ (a_{m},p_{m})\right\} _{1\leq m\leq M}$ according to the $a_m$ and use~\eqref{eq:Return_Time_Large_Deviation_Algorithm}, as described in~\ref{sec:return_time_algorithm}.
Note that in practice, with the particular choice of score function $\xi(t,x)=\obs(x)$, storing the levels $\mathcal{Q}_n^{(j)}$ for the killed trajectories directly provides the corresponding values $a_m$.

By definition, the estimators $\hat{q}_M(a)$ and $\hat{r}_M(a)$ are random variables.
In appendix~\ref{sec:AMSproperties}, we describe their statistical properties, and how to interpret them in terms of consistency and efficiency of the \ac{ams} algorithm.
In particular, we show that $\hat{q}_M(a)$ is an unbiased estimator of $q(a)$, study the variance, and show the existence of a Central Limit Theorem.

In section~\ref{sec:amsalgo_rettimes}, we proposed two choices for the number of iterations in the algorithm.
First, we described the algorithm with a fixed number of iterations $\niter$.
Alternatively, as is often seen in the \ac{ams} literature, one may decide to iterate the algorithm until all trajectories reach set $\set{B}$.
Then $\niter$ is a random number.
In that case, the threshold $a$ which defines the set $\set{B}$ becomes the control parameter for the stopping criterion.
Under those circumstances, the estimator $\hat{q}_M$ can be expressed as
\begin{equation}
  \hat{q}_M(a) = \prod_{j=1}^\niter \left( 1-\frac{\ell_j}{\nsample}\right).\label{eq:ams_estimator_q_rand}
\end{equation}
This formula remains valid in the case where the number of iterations $\niter$ is prescribed: it suffices to define the set $\set{B}$ \emph{a posteriori}, by choosing $a=\min_{1 \leq n \leq \nsample} a_n^{(J)}$ the minimum value of the $a_m$ among the final trajectories.
The formula could also be used to compute $\hat{q}_M(b)$ with $b<a$, simply by changing the number of iterations required to meet the stopping criterion.
In practice, the easiest approach is to use the expression given in~\eqref{eq:ams_estimator_q}.

In the above, we have defined the \ac{ams} estimators $\hat{q}_M$ and $\hat{r}_M$ based only on the number of trajectories generated by the algorithm.
In fact, the $\nsample$ initial trajectories and the $\tilde{\niter}$ resampled trajectories (generated during the $\niter$ iterations) are qualitatively different.
In practice, the user does not choose the parameter $M$ directly, but rather the number of ensemble members $\nsample$ on the one hand, and either the threshold $a$ or the number of iterations $\niter$ on the other hand.
As explained in appendix~\ref{sec:AMSproperties}, the number of initial trajectories $\nsample$ governs the convergence of the estimators.
Another practical constraint on the choice of $\nsample$ is the problem of \emph{extinction}: for some systems, if $\nsample$ is too small, all the members of the ensemble become identical after a number of iterations.
The other parameter (the threshold $a$ or the number of iterations $\niter$) selects the type of events we are interested in.
Indeed, from~\eqref{eq:ams_estimator_q_rand}, we obtain an approximate relation between the number of resampled trajectories $\tilde{\niter}$ and the target return times: we write $\ln \hat{q}_M(a) = \sum_{j=1}^{\niter}\ln \left(1-\frac{\ell_j}{\nsample}\right)$.
For large $\nsample$, this leads to $\ln \hat{q}_M(a) \approx - \sum_{j=1}^{\niter} \ell_{j}/\nsample \approx - \tilde{\niter}/\nsample$.
Targeting rare events with probability $10^{-\beta}$, i.e. return times of order $10^{\beta}T_a$, $\tilde{\niter}$ is then $\mathcal{O}(\nsample \beta)$.
This indicates how to choose the number of iterations $\niter$ in practice.
In particular, for rare events, we should often be in the regime $\niter=\nsample\beta$.

To sum up, to compute return time plots $r(a)$, one may either fix the target amplitude $a$, and run the algorithm for a random number of iterations, until the observable reaches $a$ for all the trajectories (i.e. until all the trajectories reach set $\set{B}$), or fix the target return time $r(a)$, and iterate the algorithm a fixed number of times by choosing $\niter=\nsample \ln(r(a)/T_a)$.
In the former case, the prescribed amplitude $a$ needs not correspond to the largest event for which we should estimate the return time, but it will approximately be the case as soon as $\nsample \ll \niter$, i.e. if $a$ is large enough for fixed $\nsample$.
Similarly, in the latter case, the largest return time computed by the algorithm will approximately be equal to the prescribed target return time when $\nsample \ll \niter$.\\

%
%

Please note that this method computes the probability to exceed a threshold $a$, by averaging over trajectories or over $K$ algorithm realisations the sampled value of $q(a)$.
This gives an unbiased estimator of $q(a)$, as explained in appendix~\ref{sec:AMSproperties}.
The standard deviation of this estimator is of order $1/\sqrt{\nexp \nsample}$.
When computing $r(a)$ through the nonlinear relation $\hat{r}(a)=-T_a/\ln\left(1-\hat{q}(a)\right)$, we thus obtain an estimator of $r(a)$ with a bias of order of $1/(\nexp \nsample)$ and a standard deviation of order $1/\sqrt{\nexp \nsample}$.
If however we had made averages over return times among algorithm realisations, then the estimator for each realisation would have been biased with a bias of order $1/\nsample$ (see appendix~\ref{sec:AMSproperties}), and the final estimator after $\nexp$ realisations would still be biased with a bias of order $1/\nsample$.

\subsection{Return times for the Ornstein--Uhlenbeck process from the \acl{tams} algorithm}
\label{sec:return_time_AMS}

We consider the Ornstein--Ulhenbeck process $X_t$ defined as
\begin{equation}
  \label{eq:OUprocess}
  dX_t=-\alpha X_t dt+\sqrt{2\epsilon}dW_t
\end{equation}
with $\alpha = 1$ and $\epsilon = 1/2$. The correlation time is $\tau_c = 1$ and the variance is $\sigma^2 = 1/2$. We now illustrate the use of the \ac{tams} algorithm for computing the return times $r(a)$ for the variable $X_t$ being larger than a threshold $a$. This amounts to choose the observable as $A(x) = x$. We use the \ac{tams} algorithm described in~\ref{sec:amsalgo_rettimes} with a score function $\xi(x,t) = x$. This choice of score function is motivated by the fact that the optimal score function is nearly independent of time, except on a small boundary layer, as explained in~\ref{sec:committor}, and that in dimension 1, the level set of $x$ will be the same as the level set of the static committor function.

The algorithm relies on three numerical parameters : the length of the generated trajectories $T_a$, the maximum threshold value $a_{max}$ and the number of replicas $\nsample$. As explained in appendix~\ref{sec:AMSproperties}, the relative error depends on $\nsample$. Additionally, one has to choose $T_a \ll \tau_c$, as explained in section~\ref{sec:return_time_rare_events}. We see empirically that a good trade-off between this requirement and computational burden is to choose trajectories of length $T_a$ equal to a few correlation times.

\begin{figure}
  \centering
  \includegraphics[width=.75\linewidth]{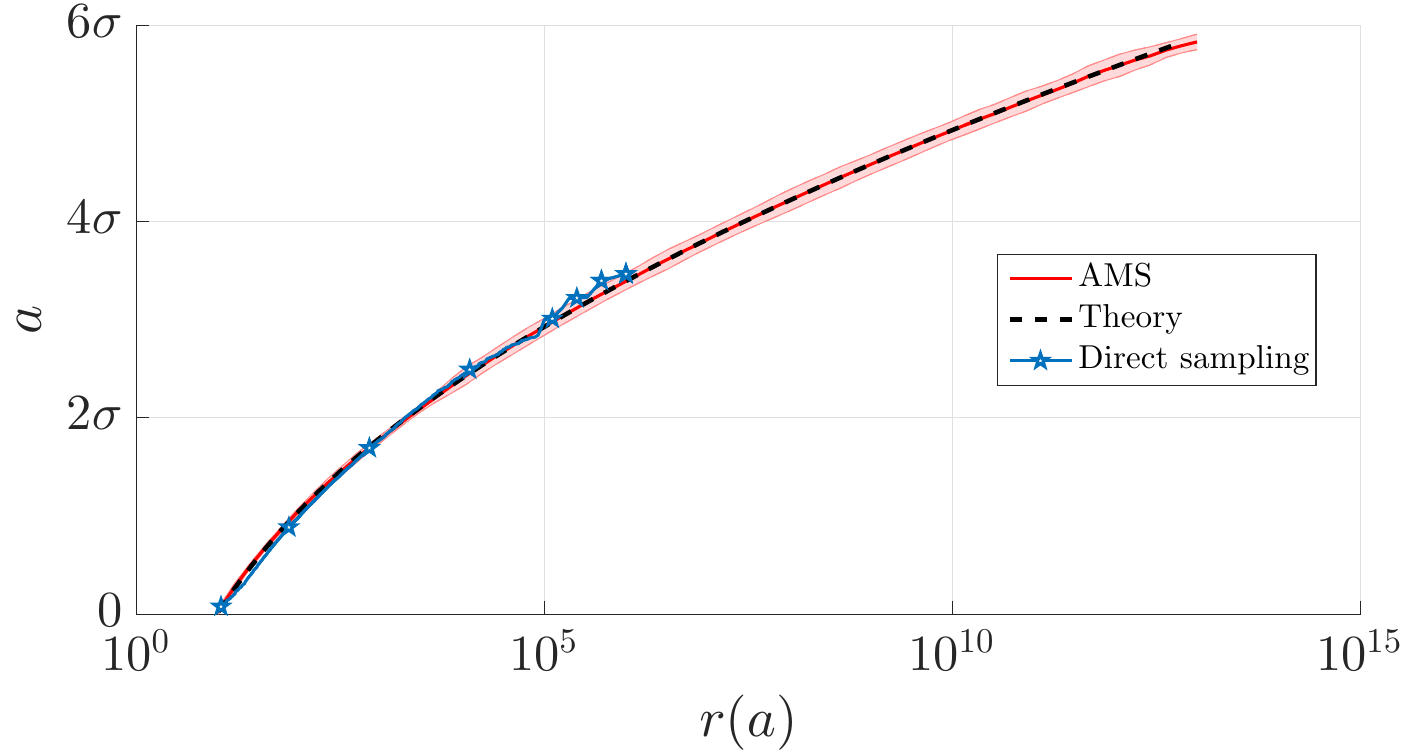}
  \caption{Return time plot for a random variable following an Ornstein--Uhlenbeck process~\eqref{eq:OUprocess} with $\alpha = 1$ and $\epsilon = 1/2$ ($\sigma=1/\sqrt{2}$ is the standard deviation). The solid red line represents the estimate obtained using the \ac{tams} with $\nsample=100$ replica, $T_a=5\tau_c$ and $a=7\sigma$. The total number of trajectories (both initial and resampled) is $M\approx 2\times 10^3$ so that the total computational cost is $\mathcal{O}(10^6 \tau_c)$. It is compared to the modified block maximum estimator $\hat{r}_B'$ applied to a sample timeseries of length $T_d = 10^6\tau_c$ (blue stars) and to the analytical result~\eqref{eq:MeanFirstPassage_S}. The shaded area represents the confidence interval on the estimation of the fluctuation amplitude $a$, for a fixed value for the return time $r(a)$. It is computed as the empirical mean over the 100 interpolated return time curves originating from the 100 independent realisations of the algorithm.}
  \label{fig:return_time_AMS}
\end{figure}

Figure~\ref{fig:return_time_AMS} shows the return time plot computed using $\nsample=100$ replicas, $T_a = 5 \tau_c$ and $a = 7\sigma$, using the \ac{tams} in conjunction with the methodology described in section~\ref{sec:ams_returntimes}.
For comparison, figure~\ref{fig:return_time_AMS} also features the theoretical value, estimated by computing the mean first-passage time (see appendix~\ref{sec:MeanFirstPassageOU}), and the estimate obtained from a direct sampling with the same computational cost as the \ac{tams} run.
We see that return times are very well recovered by the \ac{ams} algorithm.
Furthermore, figure~\ref{fig:return_time_AMS} clearly illustrates the computational gain from the \ac{tams} algorithm.
Indeed, for the same computational cost as direct sampling, the use of the \ac{tams} algorithm gives access to return times for much rarer events: we can now accurately compute return times on the order of $10^{13}$, about seven orders of magnitude larger than direct sampling.

\section{Return times sampled with the Giardina-Kurchan-Tailleur-Lecomte algorithm}
\label{sec:gktl}

In this section, we illustrate the computation of return times using the method described in section~\ref{sec:return_time_algorithm} for a \textit{time-averaged} observable.
Even though it could be done using the \ac{tams} algorithm presented in section~\ref{sec:amsalgo}, we instead illustrate the use of a different rare-event algorithm, specifically designed to compute large deviations of time-averaged dynamical observables: the \acf{gktl} algorithm~\cite{Giardina2006,Tailleur2007,Giardina2011}.

\subsection{The algorithm}
\label{sec:gktl_algo}
The underlying idea of the \acf{gktl} algorithm is to perform a biased sampling of trajectory space.
It relies on the simulation of a population of trajectories, which, unlike direct Monte-Carlo methods, interact dynamically: at regular time intervals, some members of the ensemble are killed and some are cloned according to a weight which depends on the history of the replica.
The weights are chosen such that, after several iterations of the algorithm, generated trajectories are distributed according to a probability distribution that is tilted in order to favour trajectories with large values of a chosen time averaged observable.
This sort of algorithm has first been proposed by~\cite{Giardina2006} and has been used to study rare events in both stochastic~\cite{Giardina2006,Lecomte2007b,Garrahan2007,Hurtado2009b} and deterministic systems~\cite{Giardina2006,Tailleur2007}.
The idea of sampling quantities of interest from a distribution biased in a controlled way is very general; it is referred to as \emph{importance sampling}, and was used in many different contexts (see e.g.~\cite{Berg1992,Hartmann2002} and the general references~\cite{Bucklew2004,RubinoTuffin2009}).

More precisely, we perform simulations of an ensemble of $\nsample$ trajectories $\left\{ X_{n}(t)\right\}$ (with $n=1,2,...,\nsample$) starting from random initial conditions.
Like in section~\ref{sec:AMS}, the total integration time of the trajectories is denoted $T_{a}$.
We consider an observable of interest $\obs(X(t))$ and a resampling time $\tau$.
At times $t_{i}=i\tau$ (with $i=1,2,...,T_{a}/\tau$) we assign to each trajectory $n$ a weight $W_{n}^{i}$ defined as
\begin{equation}
W_{n}^{i}=\frac{e^{k\intop_{t_{i-1}}^{t_{i}}\obs(X_{n}(t))dt}}{R_{i}}\,\,\,\mbox{with}\,\,\,R_{i}=\frac{1}{\nsample}\sum_{n=1}^{\nsample}e^{k\int_{t_{i-1}}^{t_{i}}\obs(X_{n}(t))dt}.\label{eq:Weight}
\end{equation}
For each trajectory $X_{n}$, a random number of copies of the trajectory are generated, on average proportional to the weight $W_{n}^{i}$ and such that the total number of trajectories produced at each event is equal to $\nsample$.
The parameter $k$ is chosen by the user in order to control the strength of the selection and thus to target a class of extreme events of interest.
The larger the value of $k$, the more trajectories with large values of the time average observable will survive the selection.

As mentioned above, the \ac{gktl} algorithm performs importance sampling in the space of trajectories, which is relevant for out-of-equilibrium systems.
Let us denote formally $\mathbb{P}_{0}\left(\left\{ X(t)\right\} _{0\leq t\leq T_{a}} = \left\{ x(t)\right\} _{0\leq t\leq T_{a}}\right)$ the probability to observe a trajectory $\left\{ x(t)\right\} _{0\leq t\leq T_{a}}$ in the model, and $\mathbb{P}_{k}\left(\left\{ X(t)\right\} _{0\leq t\leq T_{a}} = \left\{ x(t)\right\} _{0\leq t\leq T_{a}} \right)$ the probability to observe the same trajectory with the algorithm.
By construction of the algorithm through the weights~\eqref{eq:Weight}, we have
\begin{align}
\mathbb{P}_{k}\left(\left\{ X(t)\right\} _{0\leq t\leq T_{a}}=\left\{ x(t)\right\} _{0\leq t\leq T_{a}}\right) &\underset{\nsample\rightarrow\infty}{\sim} \frac{e^{k\int_{0}^{T_{a}}\obs(x(t))dt}}{Z(k,T_a)}\mathbb{\mathbb{P}}_{0}\left(\left\{ X(t)\right\} _{0\leq t\leq T_{a}}=\left\{ x(t)\right\} _{0\leq t\leq T_{a}}\right).\label{eq:Biased_Path_Approximation}
\end{align}
where the normalisation factor is given by $Z(k,T_a)=\mathbb{E}_{0}\left[e^{k\int_{0}^{T_{a}}\obs(X(t))dt}\right]$, denoting by $\mathbb{E}_{0}$ the expectation value with respect to $\mathbb{P}_{0}$, and $\underset{\nsample\rightarrow\infty}{\sim}$ means that this is true only asymptotically for large $\nsample$.
The typical error is of order $1/\sqrt{\nsample}$ when evaluating averages over observables.
Equation~\eqref{eq:Biased_Path_Approximation} is obtained by assuming the mean field approximation
\begin{equation}
R_{1}=\frac{1}{\nsample}\sum_{n=1}^{\nsample}e^{k\int_{0}^{t_{_{1}}}\obs(X_{n}(t))dt}\underset{\nsample\rightarrow\infty}{\sim} Z(k,t_1)= \mathbb{E}_{0}\left[e^{k\int_{0}^{t_{1}}\obs(X(t))dt}\right],\label{eq:Mean_Field_Approximation}
\end{equation}
which, by induction, and using a formula similar to~\eqref{eq:Mean_Field_Approximation} at each step of the induction, leads to~\cite{Giardina2006,Giardina2011}:
\begin{equation}
\prod_{i=1}^{T_{a}/\tau}R_{i}\underset{\nsample\rightarrow\infty}{\sim} Z(k,T_a) =\mathbb{E}_{0}\left[e^{k\int_{0}^{T_a}\obs(X(t))dt}\right].\label{eq:Estimate_Lambda}
\end{equation}
The validity of the mean field approximation and the fact that the typical relative error due to this approximation is of order $1/\sqrt{\nsample}$ has been proven~\cite{DelMoralBook,DelMoral2013} to be true for a family of rare event algorithms including the one adopted in this paper.

Formula~\eqref{eq:Biased_Path_Approximation} is valid only for times $T_{a}$ that are integer multiples of the resampling time $\tau$.
The killed trajectories have to be discarded from the statistics.
Starting from the final $\nsample$ trajectories at time $T_{a}$, one goes backwards in time through the selection events attaching to each piece of trajectory its ancestor.
In this way one obtains an effective ensemble of $\nsample$ trajectories from time 0 to time $T_{a}$, distributed according to $\mathbb{P}_{k}$.
All trajectories reconstructed in this way are real solutions of the model: we have not modified the dynamics, but only sampled trajectories according to the distribution $\mathbb{P}_{k}$ rather than according to the distribution $\mathbb{P}_{0}$.

The \ac{gktl} algorithm was initially designed to compute large deviation rate functions~\cite{Giardina2006}.
Indeed, using $\lambda(k,T_a)=\frac 1 {T_a} \ln Z(k,T_a)$, the \emph{scaled cumulant generating function}~\cite{Touchette2009} $\lambda(k)=\lim_{T_a \to +\infty} \lambda(k,T_a)$ can easily be estimated from the algorithm.
From there, the large deviation rate function $I(a)$, such that $\mathbb{P}_0\left\lbrack\int_0^{T_a} \obs(X(t))dt = T_a a\right\rbrack \asymp e^{-T_a I(a)}$, is recovered by the Legendre-Fenchel transform $I(a) = \sup_k (ka-\lambda(k))$~\cite{Touchette2009}.
In fact, the algorithm can be used to compute the statistical properties with respect to the distribution $\mathbb{P}_{0}$ of any observable, from the distribution $\mathbb{P}_{k}$.
This is done using the backward reconstructed trajectories and inverting formula~\eqref{eq:Biased_Path_Approximation}.
If, for example, one wants to estimate the expectation value of an observable $O\left(\left\{ X(t)\right\} _{0\leq t\leq T_{a}}\right)$, an estimator is given by
\begin{equation}
\mathbb{E}_{0}\left[O\left(\left\{ X(t)\right\} _{0\leq t\leq T_{a}}\right)\right]\underset{\nsample\rightarrow\infty}{\sim}\frac{1}{\nsample}\sum_{n=1}^{\nsample}O\left(\left\{ X_{n}(t)\right\} _{0\leq t\leq T_{a}}\right)\mbox{e}^{-k\int_{0}^{T_{a}}\obs(X_{n}(t))dt}\mbox{e}^{T_{a}\lambda(k,T_{a})},\label{eq:GK_O_estimator}
\end{equation}
where the $X_{n}$ are the $\nsample$ backward reconstructed trajectories.
Empirical estimators of quantities related to rare (for $\mathbb{\mathbb{P}}_{0}$) events of the kind of~\eqref{eq:GK_O_estimator} (thus using data distributed according to $\mathbb{P}_{k}$) have a dramatically lower statistical error, due to the larger number of relevant rare events present in the effective ensemble.
In particular, one can use the reconstructed trajectories to compute return times using the method described in section~\ref{sec:return_time_algorithm}.
Of course, the above formula will not perform well for quantities which are rare for the biased statistics, and we should carefully construct the effective ensemble depending on the class of observables $O$ we are trying to estimate.

\subsection{Return times for the time-averaged Ornstein--Uhlenbeck process from the \ac{gktl} algorithm}


We consider the time averaged position
\begin{equation}
  \label{eq:time_averaged}
\overline{X}_{T}(t) = \frac{1}{T}\int_{t-T}^{t}\, x(s) ds, \,\, t\in [T,T_a]
\end{equation}
where the position $x$ follows an Ornstein--Uhlenbeck process~\eqref{eq:OUprocess} between times $0$ and $T_a$.
We call $\sigma^{2}_{T}$ the variance of $\overline{X}_{T}$ and $\tau_{c,T}$ the correlation time. In this section we illustrate the application of  the \ac{gktl} algorithm to the computation of the return times $r(a)$ for $\bar{X}_T$ being larger then $a$. We make use of the \ac{gktl} algorithm with $T_a > T$, computing the time-averaged position $\bar{X}_T(t)$ for $T \leq t \leq T_a$ as a moving average.

Similarly to the case of the \ac{tams} (see section~\ref{sec:return_time_AMS}), the application of the \ac{gktl} algorithm depends on three numerical parameters: the number of trajectories $\nsample$, the length of the trajectories $T_a$ and the bias parameter $k$.
The number of trajectories $\nsample$ governs the relative error, as explained in section~\ref{sec:gktl_algo}, and one should use $T_a$ so that $T_a-T \gg \tau_{c,T}$, as explained in section~\ref{sec:return_time_rare_events}.
Finally, as for the strength of the selection $k$, its relation with the amplitude of the generated fluctuations is not known beforehand, and one has to set its value empirically~\footnote{When the duration of the average is long enough so that a \emph{large deviation regime} is attained, the relation between the value of $k$ and the typical amplitude of the fluctuations generated by the algorithm is known from the Gartner-Ellis theorem. See Ref.~\cite{Touchette2009} for further details.}.

\begin{figure}
  \centering
  \includegraphics[width=.7\linewidth]{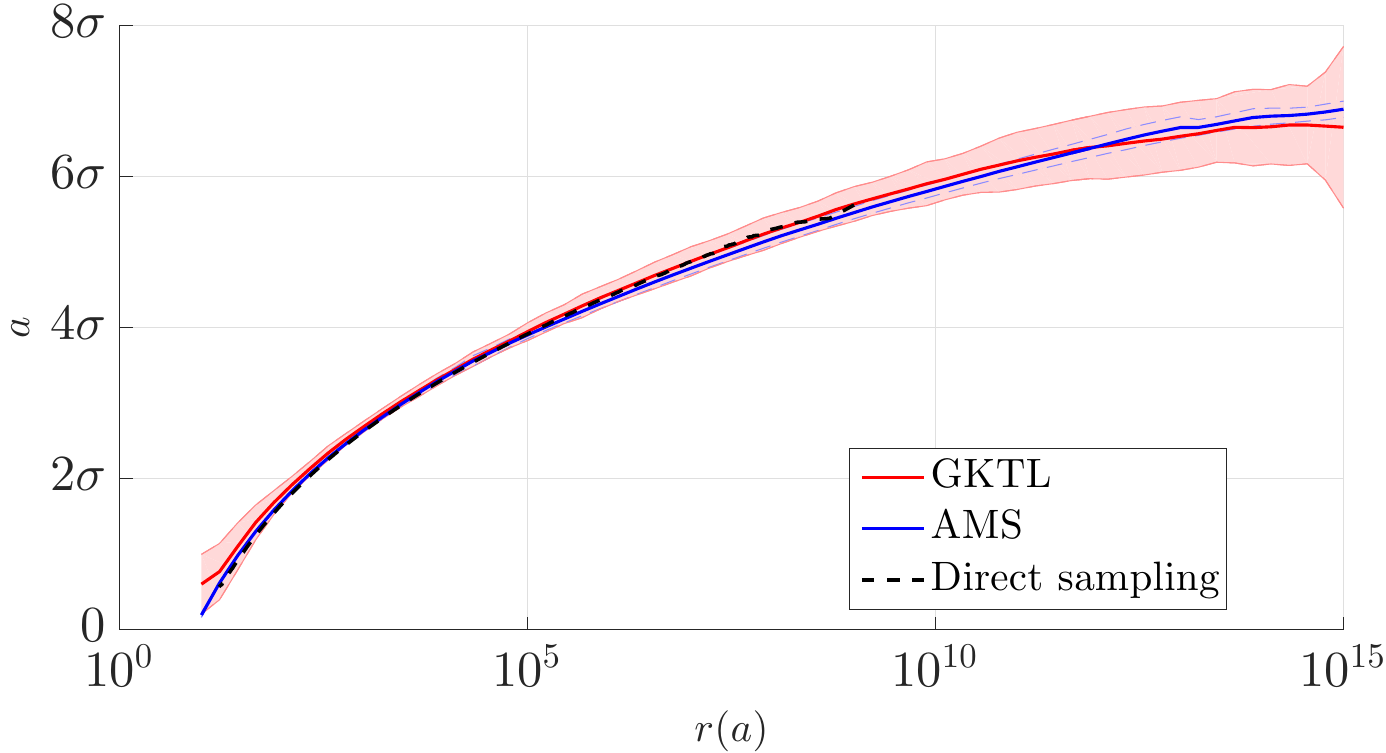}
  \caption{Return time plot for the time-averaged Ornstein--Uhlenbeck process $\overline{X}_T$~\eqref{eq:time_averaged} with $\alpha = 1$ and $\epsilon = 1/2$ ($\sigma=1/\sqrt{2}$ is the standard deviation), estimated from the \ac{gktl} algorithm (solid red line) and \ac{ams} algorithm (solid blue line). The \ac{gktl} algorithm was used with $\nsample=500$ replica, $T_a=20\tau_c$ and $k=0.9$. It was repeated $K=100$ times.
    The \ac{tams} algorithm was used with $\nsample = 100$ replicas, $T_a = 50$ and $a = 6.5\sigma_T$. It was repeated $K=10$ times.
    Finally, the dashed black line represents the result of a direct sampling over a timeseries of length $T_d = 10^9$.
    Parameters of both the \ac{gktl} and \ac{ams} algorithms were chosen so that 100 realisations of the algorithms amount to a computational cost of $\mathcal{O}(10^6 \tau_c)$. The cost of the direct sampling is $10^9 \tau_c$.The shaded area represents the confidence interval on the estimation of the fluctuation amplitude $a$, for a fixed value for the return time $r(a)$. It is computed as the empirical mean over the 100 interpolated return time curves originating from the 100 independent realisations of the algorithm.}
  \label{fig:return_time_GKTL}
\end{figure}
In Fig.~\ref{fig:return_time_GKTL}, we show the return times $r(a)$ for $\overline{X}_{T}$, with $T=10\tau_c$, computed from the \ac{gktl} algorithm described in section~\ref{sec:gktl_algo}, following the methodology described in~\ref{sec:return_time_algorithm}. 
In order to validate the computation, the estimate obtained from the algorithm is compared to the direct sampling method~\eqref{eq:Return_Times_Rare-1}.
For rare events ($r(a) \gg \tau_{c,T}$), the results from the \ac{gktl} algorithm agree well with direct sampling.
Furthermore, the comparison of the computational costs for the two different methods shows the efficiency of the algorithm.
Indeed, for direct sampling, the length of the sample trajectory, $10^9 \tau_c$ in the case of Fig.~\ref{fig:return_time_GKTL}, naturally sets an upper bound on the return times one is able to compute.
By contrast, the total cost of the \ac{gktl} estimate is $10^6 \tau_c$ and one can see in Fig.~\ref{fig:return_time_GKTL} that it allows to reach return times larger by many orders of magnitude.
\begin{figure}
  \centering
  \includegraphics[width=.7\linewidth]{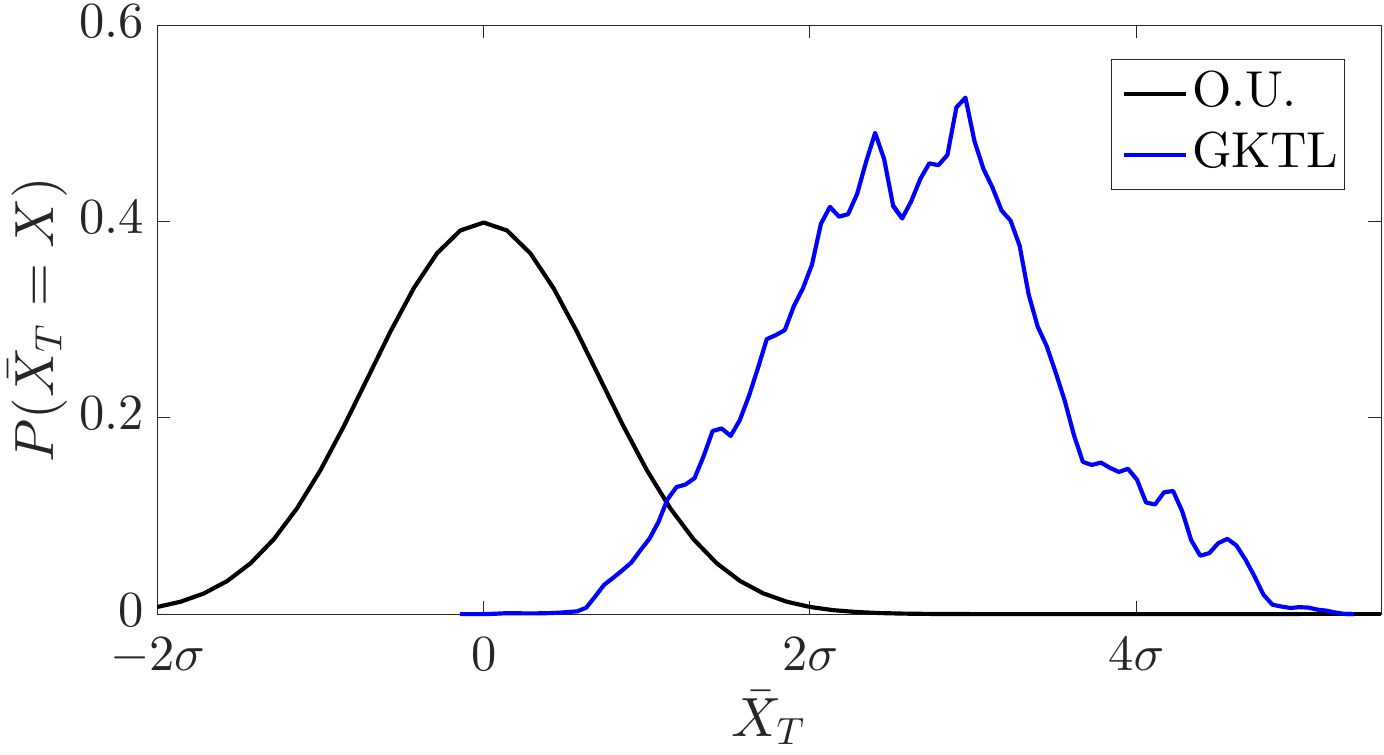}
  \caption{\ac{pdf} of the time-averaged observable $\bar{X}_T$, with $T = 10\tau_c$, for the Ornstein--Uhlenbeck process with $\alpha=1$ and $\epsilon=1/2$ ($\sigma=1/\sqrt{2}$ is the standard deviation): computed from a direct simulation of length $T_d = 10^6$ (black curve), and based on the trajectories generated by the \ac{gktl} algorithm with $500$ replicas, $T_a=20\tau_c$ and $k=0.9$ (blue curve).}
  \label{fig:pdf_GKTL}
\end{figure}
Figure~\ref{fig:pdf_GKTL} shows an estimate of the \ac{pdf} for $\bar{X}_T$ along the trajectories generated using the \ac{gktl} algorithm.
Even though importance sampling is performed for the observable $\bar{X}_{T_a}$, the observable averaged over the whole trajectory of length $T_a$, it better samples the tail of the \ac{pdf} for $\bar{X}_T$, resulting in better estimation of the corresponding return times.

Figure~\ref{fig:return_time_GKTL} also shows the return time for $\bar{X}_T(t)$ being larger than $a$ computed using the \ac{tams} algorithm (see section~\ref{sec:amsalgo_rettimes}). We use as a score function the time-averaged observable itself $\xi(t) = \bar{X}_T(t)$, for $T\leq t \leq T_a$. The selection is then done according to the maximum value of $\bar{X}_T(t)$ for each trajectory for $T\leq t \leq T_a$. More precisely, following the notations of section~\ref{sec:amsalgo_rettimes}, for iteration $j$ we denote $\mathcal{Q}_j^\star$ the lowest maximum of $\bar{X}_T$ over the trajectories in the set $\{x_n^{(j)}(t)\}_{0\leq t \leq T_a, 1\leq n \leq N}$. Following the \ac{tams} algorithm described in section~\ref{sec:amsalgo_rettimes}, the $l_j$ new replica are defined by copying the trajectories $x_{n_\ell}^{(j-1)}$ from $0$ to the smallest time $t$ such that $\bar{X}_{T,n_\ell}^{(j-1)}(t) > \mathcal{Q}_j^\star$ and simulating the rest of the trajectory from this time to $T_a$.

The agreement between the two estimates illustrates that the method to compute return times from rare event algorithms proposed in~\ref{sec:return_time_algorithm} can be applied to any rare event algorithm suitable for the type of observable under study.
Here, while the \ac{ams} algorithm allows for computing return times for both the instantaneous and time-averaged observables, the \ac{gktl} algorithm is not suited for instantaneous observables.

\section{Application: Extreme drag force on an object immersed in a turbulent flow}
\label{sec:applications}

A key issue with rare event algorithms is to understand if they are actually useful to compute rare events and their probabilities for actual complex dynamical systems.
The \ac{ams} algorithm has shown to be very efficient for partial differential equations with noise~\cite{Rolland2016}.
In this section, we give a brief illustration that more complex dynamics can be studied.
We illustrate the computation of return times using rare event algorithms for a turbulent flow.
The possible limitations of rare event algorithms are further discussed in the conclusion.

Unlike simple low-dimensional models, such as the Ornstein-Uhlenbeck process, numerical simulations of turbulent flows of interest for physicists and/or engineers require tremendous computational efforts.
As a consequence, direct sampling of rare events based on a long time series is simply unthinkable for such systems.
A common practice in the engineering community is to generate synthetic turbulent flows, without resolving explicitly the small scales, to study numerically the physical phenomena of interest~\cite{Spalart2000,Moin2002}.
However, the main difficulty is to capture synthetically the correct long-range (spatio-temporal) correlations of turbulence and such approaches can not capture the essential effects of coherent structures.
We show here that rare event methods such as the \ac{gktl} and the \ac{ams} algorithms can be used in order to study extremes in turbulent flows without having to rely on such modelling.

The example we consider is the sampling of extreme fluctuations of the mechanical stresses caused by a turbulent flow on an immersed object.
Being able to compute flow trajectories associated to such extremes is of great interest both for fundamental issues and applied problems, such as reliability assessment for industrial structures.
More specifically, we focus here on the averaged drag $F_{T}(t) = \frac{1}{T}\int_{t}^{t+T}f_d(t')\mathrm{d}t'$, which corresponds to the averaged sum of the efforts from the flow, projected along the flow direction.
The length of the averaging window depends on the nature of the application.
For instance, it could be related to the typical response time of a material, in order to average out high frequency excitation that has a minor impact on the deformation of the structure.
Note that the choice of the observable is arbitrary and one could choose to study other related physical quantities, such as the lift or torque.

\begin{figure}[h]
  \centering
  \includegraphics[width=.8\textwidth]{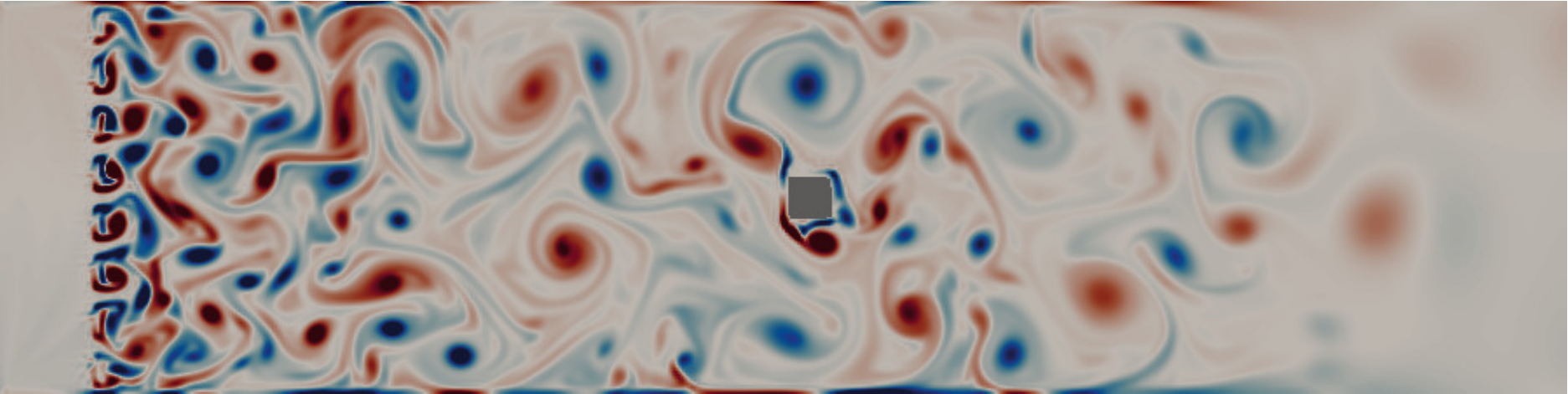}
  \caption{Snapshot of a typical vorticity field of the flow under study. A steady parabolic velocity profile is imposed at the inlet. Turbulence is then generated by a grid. We used the \ac{gktl} algorithm the compute the return times of the average drag over the square here marked by the grey area.}\label{fig:snapshots}
\end{figure}

In order to provide a \textit{proof-of-concept} for such rare events approaches for turbulent flows, we compute the return time for extreme values of the drag in a simple academic flow.
The setup we consider, illustrated in Fig.~\ref{fig:snapshots}, is that of a two-dimensional channel flow, with a square obstacle immersed in the middle of the domain.
Turbulence is generated upstream by means of a grid.
This flow is simple enough so that long time series can be obtained in a reasonable amount of computational time, allowing for the computation of reference return times.
In practice, we carry out a direct numerical simulation using the Lattice Boltzmann Method~\cite{Chen1998c}, which offers low implementation effort for performances comparable to other methods for such simple geometries and boundary conditions.
The application of the \ac{gktl} and \ac{ams} algorithms to deterministic dynamics requires that some randomness is artificially introduced in the dynamics so that copies originated from the same parent follow different paths.
This can be achieved by randomly perturbing the restart state at branching points.

\begin{figure}[thb]
  \centering
  \includegraphics[width=.6\textwidth]{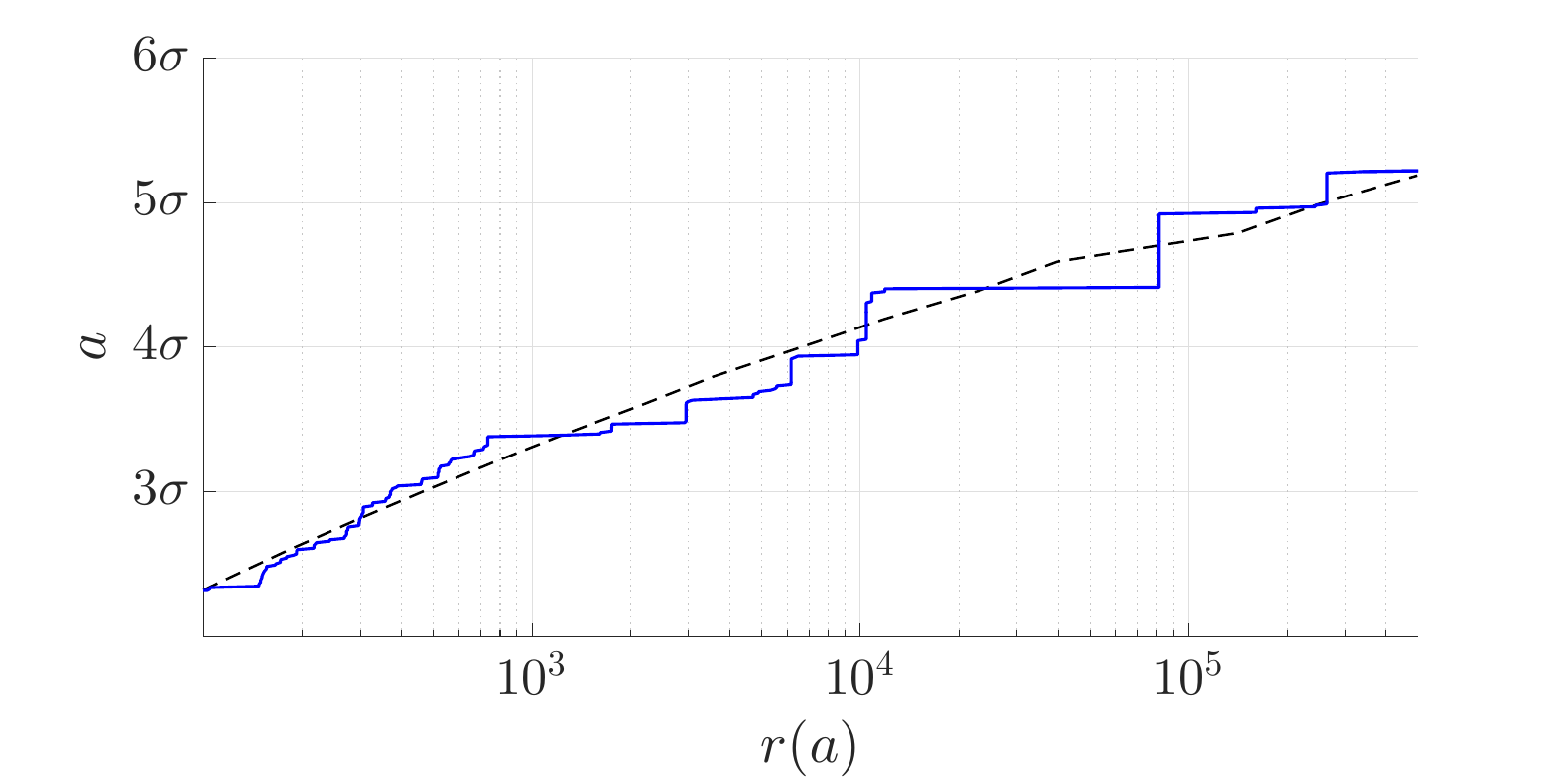}
  \caption{Illustration of the computation of return times for the averaged drag over the square obstacle pictured in Fig.~\ref{fig:snapshots}. The averaging window is $5$ correlation times. The dashed black line represents the reference return times computed from a timeseries spanning $10^6$ correlation times, using~\eqref{eq:Return_Times_Rare-1}. The solid blue line represents the return times obtained using the \ac{gktl} algorithm.}
  \label{fig:return_time_drag}
\end{figure}
Figure~\ref{fig:return_time_drag} illustrates the computation of the return times for the drag averaged over $5$ correlation times using the \ac{gktl} algorithm.
It shows that the use of the algorithm makes accessible the computation of rare events at a much lower computational cost than direct sampling.
More precisely, the algorithm was applied using $\nsample=128$ replicas simulated over $10$ correlation times.
The return time curve presented in Fig.~\ref{fig:return_time_drag} is based on the data from $\nexp=10$ repetitions of the algorithm, leading to an overall computational cost of, roughly,  $10^4$ correlation times.
From a direct sampling of similar computational cost, the rarest accessible event has a return time close to the computational cost itself, in this case is $10^4$. Figure~\ref{fig:return_time_drag} shows that the use of the GKTL algorithm allows for the computation of return times of much rarer events.
The reference curve was computed from a time series spanning $10^6$ correlation times.
For events having a return time close to $5\cdot 10^5$ correlation times, the computational cost of estimating the return times using the \ac{gktl} algorithm is $50$ times lower than direct sampling.

\begin{figure}[h]
  \centering
  \includegraphics[width=.6\textwidth]{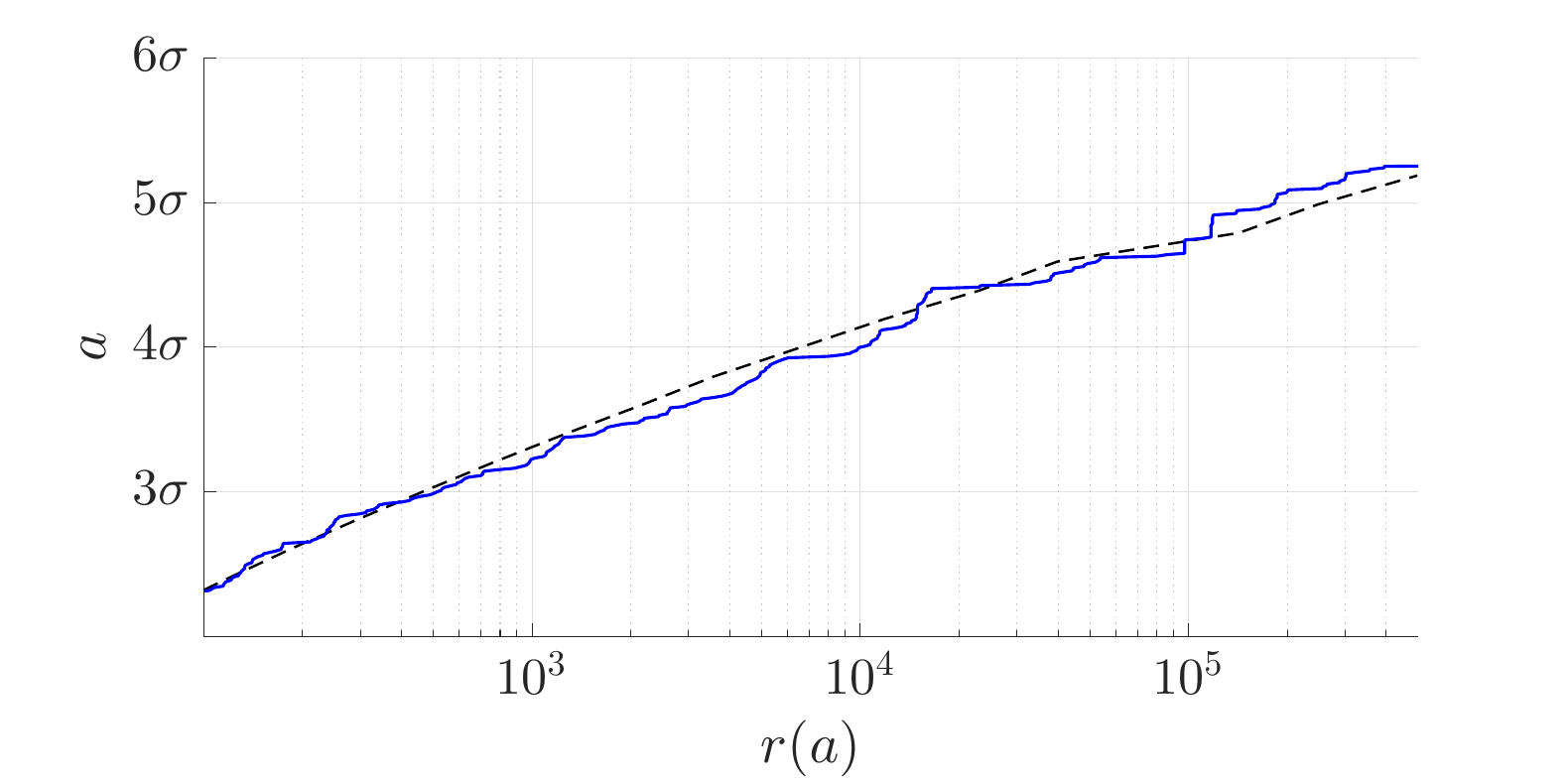}
  \caption{Illustration of the computation of return times for the averaged drag over the square obstacle pictured in Fig.~\ref{fig:snapshots}, using $50$ repetitions of the \ac{gktl} algorithm. The parameters are the same as in Fig.~\ref{fig:return_time_drag}. This figure illustrates the reduction in the occurrence of plateaus for the return time curve obtained using the \ac{gktl} algorithm. The dashed black line represents the reference return times. The solid blue line represents the return times obtained using the \ac{gktl} algorithm.}
  \label{fig:return_time_drag_50reps}
\end{figure}

The occurrence of plateaus in Fig.~\ref{fig:return_time_drag} is due to the increasing multiplicity of trajectories as the amplitude $a$ increases.
Indeed, because of the selection procedure involved in the \ac{gktl} algorithm, a subset of trajectories can share the same ancestor.
Henceforth, they are likely to differ only by a small time-interval at the end of their whole duration. In such cases, it is common that the maximum over the trajectory is attained in earlier times.
As a consequence, this subset of trajectories will contribute the same value to the set of maxima from which return times are computed.
This effect is accentuated in the present case of a deterministic system, as it takes some time for copies to separate after being perturbed at a branching point.
A straightforward way of mitigating the occurrence of such plateaus is to increase the number of trajectories or/and the number of repetitions of the algorithm.
As an illustration, Fig.~\ref{fig:return_time_drag_50reps} shows the return time plot obtained using $50$ repetitions instead of $10$ in Fig.~\ref{fig:return_time_drag}.

\section{Conclusion}
\label{sec:cl}

In this paper, we have considered the question of estimating the return time of rare events in dynamical systems.
We have compared several estimators, using both usual timeseries (generated with direct numerical simulations) and rare event algorithms, by generalising the approach relating the return times to the extrema over trajectory blocks.
This approach relies on the fact that rare events behave, to a good approximation, like a Poisson process: this allows for the derivation of a simple formula (see~\eqref{eq:Return_Times_Rare}) for estimating the return times based on block maxima.
We slightly improved this formula (see~\eqref{eq:Return_Times_Rare-1}), and further showed that it was possible, provided only minor modifications, to evaluate it with data produced by rare event algorithms.
Indeed, while the traditional block maximum method consists in dividing a given trajectory in blocks with arbitrary length (larger than the correlation time of the system, and smaller than the return time one seeks to estimate), there is a class of rare event algorithms which yields precisely an ensemble of trajectories exhibiting the rare event more often than direct simulation, together with the probability of observing each member of the ensemble.
Hence, we have generalised the block maximum formula to non-equiprobable trajectory blocks; this allowed us to use directly rare event algorithms, such as the \ac{ams} and the \ac{gktl} algorithm, to estimate return times for rare events.
Using the Ornstein--Uhlenbeck process as an illustration, we showed that the method is easy to use and accurately computes return times in a computationally efficient manner.
Indeed, compared to direct sampling, combining the generalised block maximum approach to rare event algorithms allowed for computing return times many orders of magnitude larger, at fixed computational cost.
This method does not depend on the dynamics of the system or on the type of observable, as long as a suitable rare event algorithm is selected.
As an illustration, we computed return time plots for both instantaneous and time-average observables for the Ornstein--Uhlenbeck process, using the \ac{ams} and the \ac{gktl} algorithms.
This approach paves the way to numerical computation of return times in complex dynamical systems.
To showcase the potential of the method, we discussed briefly an application of practical interest: extreme values of the drag force on an object immersed in turbulent flows.
Another example of application given very recently is the study of heat waves~\cite{Ragone2017}.

A key issue with rare event algorithms is to understand if they are actually useful to compute rare events and their probabilities for actual complex dynamical systems.
Many of the proposed  approaches fail to pass such a test, either because the algorithm is too complex to be used for complex dynamical systems, or the algorithm is restricted to specific systems (equilibrium or reversible dynamics, diffusions with small noises), or the algorithm simply fails.
A key issue with many potentially successful rare event algorithms, for instance the \ac{ams} algorithm and the \ac{gktl} algorithm among others, is that their success depends much on the quality of the rule used for selecting trajectories.
For instance the \ac{ams} or the \ac{tams} algorithm rely on a score function, and the \ac{gktl} use as a selection rule the increment of a the time average which one aims at computing.
Whenever one uses a good score function, those algorithms are extremely useful and show tremendous sampling improvements~\cite{Rolland2016}.
For the \ac{ams} algorithm, the choice of a good score function often relies on a good rough qualitative understanding by the user of the effective dynamics that leads to the rare events.
Then the \ac{ams} algorithm leads to excellent quantitative results, even with complex dynamical systems (see for instance ~\cite{Rolland2016}).
Several examples have illustrated than those algorithm may fail to lead to improvement in other cases, see for instance~\cite{nemoto2016population}.
Faced with such difficulties, one may either use an empirical approach, or try to improve the algorithms in order to cure potential problems, as we explain now.

The empirical approach consists in identifying a priori the conditions for success of the algorithms and identify relevant dynamical phenomena that fulfil these conditions.
For the \ac{ams} algorithm this amounts to understanding sufficiently well the dynamics, in order to be able to define a macroscopic variable that will describe well the dynamics leading to the extremes, and to propose a related score function.
The algorithm may also be used to test some hypothesis on such macroscopic variables, and learn about the dynamics.
The \ac{gktl} algorithm is usually successful in conditions when the sampling of time averages is dominated by a persistent macroscopic state.

Several authors have proposed new algorithms to cure some of the problems.
A class of algorithms seek at changing the dynamics such that the computation will be more efficient (see for instance ~\cite{vanden2012rare} for diffusions with small noise, or~\cite{nemoto2016population} in relation with the \ac{gktl} algorithm and references therein).
Those methods are limited to diffusions, as they require to relate the statistics of paths for different dynamics, for instance through the Girsanov formula.
They can involve recursive learning of an optimal dynamics and be very successful for dynamics with a few degrees of freedom~\cite{nemoto2016population}.
Another class of algorithms, milestoning (see~\cite{faradjian2004}), is aimed at computing a reduced description of the original dynamics, that can afterwards permits to efficiently compute dynamical quantities, for instance first passage times (see~\cite{schutte2011} and references therein).

\begin{acknowledgments}
  The research leading to these results has received funding from the European Research Council under the European Union's seventh Framework Program (FP7/2007-2013 Grant Agreement No. 616811).
  This project has received funding from the European Union's Horizon 2020 research and innovation programme under the Marie Sk\l odowska-Curie grant agreement No 753021.
  We would like to thank Emmanuel L\'ev\^eque for insightful discussions, and Joran Rolland and two anonymous referees for their comments which helped improving the manuscript.
  Simulations have been performed on the local HPC facilities at Ecole Normale Sup\'erieure de Lyon (PSMN) and Ecole centrale de Lyon (PMCS2I).
  These HPC facilities are supported by the Auvergne-Rh\^one-Alpes region (GRANT CPRT07-13 CIRA) and the national Equip@Meso grant (ANR-10-EQPX-29-01).
\end{acknowledgments}

\appendix

\section{Mean first-passage time for the Ornstein--Uhlenbeck process\label{sec:MeanFirstPassageOU}}

Throughout the paper, we consider as an example the Ornstein--Uhlenbeck process:
\begin{equation}
dX_{t}=b(X_{t})dt+\sqrt{2\epsilon}dW_{t},\label{eq:OUprocess_appendix}
\end{equation}
where $W_{t}$ is the standard Wiener process and the drift term is linear: $b(x)=-\alpha x$.
We write the corresponding Fokker-Planck equation for the probability density $P(x,t)$ of the random variable $X_t$:
\begin{equation}
  \frac{\partial P}{\partial t}=LP, \text{ with } LP=-\frac{\partial[b(x)P(x,t)]}{\partial x}+\epsilon\frac{\partial P(x,t)}{\partial x^{2}}.
\end{equation}
The stationary probability density is $P_{s}(x)=\sqrt{\frac{\alpha}{2\pi\epsilon}}e^{-\frac{\alpha x^{2}}{2\epsilon}}$: $LP_s=0$.
We shall denote the standard deviation by $\sigma=\sqrt{\epsilon/\alpha}$.

For a threshold value $a$ much larger than the standard deviation ($a\gg\sqrt{\epsilon/\alpha}$), the return time $r(a)$ should be well approximated by the mean first-passage time $\mathbb{E}[\tau_{a}]$, where $\tau_{a}=\min\{t\geq0|X_{t}\geq a\}$.
Computing the mean first-passage time for such a simple stochastic process is a classical textbook problem (see for instance~\cite[\S~5.5]{GardinerBook}): we consider the transition probability $P(x',t|x,0)$, which also satisfies the Fokker-Planck equation, with initial condition $P(x',0|x,0)=\delta(x'-x)$.

We now introduce the quantity $G(x,t)=\int_{-\infty}^{a}dx'P(x',t|x,0)$, with the initial condition $G(x,0)=\chi_{]-\infty,a[}(x)$, where $\chi$ is the indicator function, and with absorbing boundary conditions at $a$, $G(a,t)=0$.
Using the backwards Kolmogorov equation for the transition probability: $\partial_t P(x',0|x,t)=L^\dagger P(x',0|x,t)$, and using time-homogeneity $P(x',t|x,0)=P(x',0|x,-t)$, we see that the evolution of $G$ is also governed by $\partial_{t}G=L^{\dagger}G$.
$G(x,t)$ is the probability that a particle initially at position $x$ has not reached $a$ after time $t$.
In other words, it is the probability, conditioned on the initial condition $x$, that $\tau_{a}>t$.
The moments of the first-passage time follow directly:
\begin{equation}
\mathbb{E}_{x}[\tau_{a}^{n}] = -\int_0^{+\infty} t^n \partial_t G(x,t) dt  = n\int_{0}^{+\infty}t^{n-1}G(x,t)dt.
\end{equation}
From there, a recursion relation can be obtained for the moments of $\tau_{a}$:
\begin{equation}
  \mathbb{E}_{x}[\tau_{a}^{n}] = - \frac{1}{n+1} \left\lbrack b(x)\frac{\partial}{\partial x} + \epsilon \frac{\partial^2}{\partial x^2} \right\rbrack \mathbb{E}_{x}[\tau_{a}^{n+1}].
\end{equation}
In particular, with $\mathbb{E}_{x}[\tau_{a}^{0}]=1$, we obtain an exact formula for $\mathbb{E}_{x}[\tau_{a}]$:
\begin{align}
\mathbb{E}_{x}[\tau_{a}] & =\frac{1}{\epsilon}\int_{x}^{a}dye^{\frac{\alpha y^{2}}{2\epsilon}}\int_{-\infty}^{y}dze^{-\frac{\alpha z^{2}}{2\epsilon}}=\frac{\pi}{\alpha}\left\{ \text{{erfi}}\left(\sqrt{\frac{\alpha}{2\epsilon}}a\right)-\text{{erfi}}\left(\sqrt{\frac{\alpha}{2\epsilon}}x\right)\right\} -\frac{\sqrt{\pi}}{\alpha}\int_{\sqrt{\frac{\alpha}{2\epsilon}}x}^{\sqrt{\frac{\alpha}{2\epsilon}}a}due^{u^{2}}\text{{erfc}}(u),\label{eq:MeanFirstPassage_x0}
\end{align}
when $x<a$, and 0 otherwise, where $\text{erfc}$ and $\text{erfi}$ are the complementary and imaginary error functions, respectively~\cite{AbramowitzBook}.
It is straightforward to obtain the mean first passage time conditioned on the stationary measure:
\begin{equation}
\mathbb{E}_{s}[\tau_{a}]=\int_{-\infty}^{+\infty}dxP_{s}(x)\mathbb{E}_{x}[\tau_{a}]=\sqrt{\frac{\alpha}{2\pi\epsilon^{3}}}\int_{-\infty}^{a}dye^{\frac{\alpha y^{2}}{2\epsilon}}\left(\int_{-\infty}^{y}dze^{-\frac{\alpha z^{2}}{2\epsilon}}\right)^{2}.\label{eq:MeanFirstPassage_S}
\end{equation}
The above formula provides the theoretical prediction against which numerical estimates of return times for the Ornstein--Uhlenbeck process are compared in the paper.

\section{Statistical properties of AMS estimators}\label{sec:AMSproperties}

The standard way of analysing the efficiency of an estimator $\hat{\theta}_\nsample$ (or rather, a family of estimators indexed by a parameter $\nsample$, e.g. a sample size) of a quantity $\theta$ is to consider the mean-square error:
\begin{equation} \text{MSE}(\nsample)=\mathbb{E}|\hat{\theta}_\nsample-\theta|^2=\bigl(\mathbb{E}[\hat{\theta}_\nsample]-\theta\bigr)^2+\text{Var}\bigl(\hat{\theta}_\nsample\bigr),
\end{equation}
which is decomposed into the contributions of the bias $b(\nsample)=\mathbb{E}[\hat{\theta}_\nsample]-\theta$ (which represents the systematic, or model, error) and of the variance $\text{Var}\bigl(\hat{\theta}_\nsample\bigr)$ (which represents the statistical error).
For some error tolerance $\epsilon>0$, the cost of the simulation is the expected cost of one realisation of the algorithm using a parameter $\nsample$ such that $\text{MSE}(\nsample)\le \epsilon^2$: finding optimal $\nsample$ requires a bias-variance trade-off.
The precision of the estimation is improved by controlling the bias, and the fluctuations by controlling the variance.

We now consider the \ac{ams} estimator $\hat{q}_\nsample(a)$, defined in section~\ref{sec:ams_returntimes}.
Note that we index this estimator with the number of initial trajectories $\nsample$: as we shall see, this is the parameter which controls the statistical properties, and not the total number of sampled trajectories $M=\nsample+\tilde{\niter}$.
One of the main properties of the \ac{ams} algorithm is the following unbiasedness result, see~\cite{Brehier2016a} for more general statements, and discussion on the influence of the time discretization of the Markov dynamics.
\begin{theo}
  For every $\nsample$, for every score function $\xi$, $\hat{q}_\nsample$ is an unbiased estimator of $q$:
  \begin{equation}
    \mathbb{E}[\hat{q}_\nsample]=q.
  \end{equation}
\end{theo}
Thus only the statistical error $\text{Var}(\hat{q}_\nsample)$ depends on the choice of $\nsample$, and, more importantly, on the score function $\xi$; see~\cite{Brehier2016a,Rolland2015} for extensive numerical simulations concerning the role of the score function.
In practice, it is recommended in~\cite{Brehier2016a} that one computes empirical averages $\overline{q}_{\nsample,\nexp}=\frac{1}{\nexp}\sum_{k=1}^{\nexp}\hat{q}_\nsample^{(k)}$ over $\nexp$ independent realisations of the algorithm, with large $\nexp$: the associated mean-square error is $\text{MSE(\nsample,\nexp)}=\frac{\text{Var}(\hat{q}_\nsample)}{\nexp}$.
Moreover, repeating the experience with different choices of score functions is a way to validate the results, checking the overlap of confidence intervals.

In addition, it has been proved, in different contexts, see~\cite{Brehier2015b,Cerou2016}, that $\hat{q}_\nsample$ is a consistent estimator of $q$: the convergence $\hat{q}_\nsample\underset{\nsample\to \infty}\to q$ holds true, in probability.
More precisely, it is proved in~\cite{Cerou2016}, that the estimator $\hat{q}_\nsample$ satisfies a Central Limit Theorem,
\begin{equation}
  \sqrt{\nsample}\bigl(\hat{q}_\nsample-q\bigr)\underset{\nsample\to \infty}\to\mathcal{N}(0,\sigma^2(\xi,q)),
\end{equation}
with an asymptotic variance $\sigma^2(\xi,q)\in [-q^2\ln q,2q(1-q)]$.
The minimal variance $-q^2\ln q$ is obtained when choosing
\begin{equation}
  \xi(y)=\overline{\xi}(y) \equiv \mathbb{P}_{y}(\tau_\set{B}<\tau_\set{A}).
\end{equation}
In practice, the optimal score function $\overline{\xi}$, also referred to as the \emph{committor}, is of course not known; note that the estimated probability satisfies $q=\xi(y_0)$.
Below we will discuss more precisely the statistical properties of the estimators $\hat{q}_\nsample$ and $\hat{r}_\nsample$ when choosing $\overline{\xi}$ as the score function.

Note that $\sigma^2(\xi,q)\le 2q(1-q)=2\text{Var}(\mathcal{P})$, where $\mathcal{P}$ is a Bernoulli random variable with mean $q$.
This ensures that in terms of variance the \ac{ams} algorithm performs better than or similarly to the crude Monte Carlo method, in the rare event regime $q\to 0$; moreover, the \ac{ams} algorithm with optimal score function outperforms the crude Monte-Carlo method (please note that this is the variance normalised by $\nsample$, where $\nsample$ is the number of initial trajectories, and not by $M$, where $M=\nsample+\tilde{\niter}$ is the total number of computed trajectories).

Note that $\mathbb{E}[\hat{r}_\nsample]\neq r$, and thus $\hat{r}_\nsample$ is not an unbiased estimator of $r$.
However, a Central Limit Theorem still holds true: since $\hat{r}_\nsample=\phi(\hat{q}_\nsample)$ and $r=\phi(q)$ for some function $\phi$, such that $\phi'(q)\neq 0$, the $\delta$-method~\cite{vanderVaart1998} implies
\begin{equation}
  \sqrt{\nsample}\bigl(\hat{r}_\nsample-r\bigr)\underset{\nsample\to \infty}\to \mathcal{N}\bigl(0,\sigma^2(\xi,q)(\phi'(q))^2\bigr),
\end{equation}
where $q^2\phi'(q)\underset{q\to 0}\to T_a$, with $T_a$ the size of the window.
The estimators for the return time $r$ correspond to the choices $\phi(q)=-T_a/\ln(1-q)$ or $\phi(q)=T_a/q$.

For an arbitrary choice of the score function $\xi$, it is not possible in general to obtain precise results concerning the bias for the return time $\hat{r}_\nsample$, and the asymptotic variance.
However, when the optimal score function $\overline{\xi}(y)=\mathbb{P}_y(\tau_\set{B}<\tau_\set{A})$ is used, elementary arguments are sufficient to analyse the statistical properties of estimators $\hat{q}_\nsample$ and $\hat{r}_\nsample=\frac{1}{\hat{q}_\nsample}$ (with $T_a=1$).
The key property~\cite{Brehier2015b,Guyader2011,Simonnet2016}, is that, when using the optimal score function, the number of iterations $\niter$ follows a Poisson distribution, with parameter $-\nsample\ln q$.
This situation is referred to as the idealised case in the mathematical literature.
Since $\hat{q}_\nsample=\bigl(1-\frac{1}{\nsample}\bigr)^{\niter}$, proving the following results is straightforward: first, concerning the bias,
\begin{equation}
  \mathbb{E}\lbrack\hat{q}_\nsample\rbrack=q,\qquad \mathbb{E}\left\lbrack\frac{1}{\hat{q}_\nsample}\right\rbrack-\frac{1}{q}\underset{\nsample\to\infty}\sim \frac{-\ln q }{q\nsample}.
\end{equation}
Second, concerning the asymptotic variance,
\begin{equation}
  \text{Var}(\hat{q}_\nsample) = q^2\bigl(q^{-\frac{1}{\nsample}}-1\bigr)\underset{\nsample\to \infty}\sim \frac{-q^2\ln q }{\nsample},\qquad \text{Var}\left(\frac{1}{\hat{q}_\nsample}\right)\underset{\nsample\to \infty}\sim \frac{-\ln q}{\nsample q^2}.
\end{equation}
Note that relative bias and variance are both of size $\frac{-\ln q}{\nsample}$.
The derivation of the Central Limit Theorem~\cite{Brehier2016b}, and Large Deviations results~\cite{Brehier2015a} is also straightforward in the idealised case.

\bibliographystyle{apsrev4-1}
\bibliography{returntimes}

\end{document}